\documentclass[nofootinbib,preprint,preprintnumbers,a4paper,11pt]{revtex4}
\usepackage{epsfig}
\usepackage{footnote}
\usepackage{ulem}
\usepackage{color}
\usepackage{array}
\usepackage{amssymb}
\usepackage{amsmath}
\usepackage{graphicx}
\usepackage{epstopdf}
\usepackage{subfigure}
\usepackage{longtable}
\usepackage{verbatim}
\usepackage{amsfonts}
\usepackage{hyperref}
\usepackage{makecell}

\begin{document}

\title{Charmed baryon decays in $SU(3)_F$ symmetry}

\author{Cai-Ping Jia$^{1}$\footnote{Email: jiacp17@lzu.edu.cn}, Di Wang$^{1,2}$\footnote{Email: dwang15@lzu.edu.cn, corresponding author} and Fu-Sheng Yu$^{1}$\footnote{Email: yufsh@lzu.edu.cn, corresponding author}}
\address{%
$^1$ School of Nuclear Science and Technology,  Lanzhou University,  Lanzhou 730000,  People's Republic of China \\
$^{2}$ IPPP, Department of Physics, University of Durham,
DH1 3LE, United Kingdom
}

\begin{abstract}

In the recent years, fruitful results on charmed baryons are obtained by BESIII, Belle and LHCb.
We investigate the two-body non-leptonic decays of charmed baryons based on the exact flavor $SU(3)$ symmetry without any other approximation.
Hundreds of amplitude relations are clearly provided, and are  classified according to the $I$-, $U$- and $V$-spin symmetries.
Among them, some amplitude relations are tested by the experimental data, or used to predict the branching fractions based on the exact flavor symmetry without any other approximation.
Some relations of $K^0_S-K^0_L$ asymmetries and $CP$ asymmetries are obtained under the $U$-spin symmetry in the modes of charmed baryon decaying into neutral kaons.
Besides, the $U$-spin breaking effect is explored in the $\Lambda_c^+\to \Sigma^+K^{*0}$ and $\Xi_c^+\to p\overline{K}^{*0}$ modes.

\end{abstract}

\maketitle

\section{Introduction}

Charmed baryon decays have attracted great attentions recently.
Many new measurements were performed by the BESIII \cite{Ablikim:2019zwe,Ablikim:2018czr,Ablikim:2018byv,Ablikim:2018jfs,Ablikim:2018woi,Ablikim:2018bir,Ablikim:2015prg,Ablikim:2015flg,Ablikim:2016tze,Ablikim:2016mcr,Ablikim:2016vqd,Ablikim:2017ors,Ablikim:2017iqd},
Belle \cite{Li:2019atu,Li:2018qak,Berger:2018pli,Zupanc:2013iki,Yang:2015ytm,Pal:2017ypp}, and
LHCb \cite{Aaij:2019lwg,Aaij:2019kss,Aaij:2018dso,Aaij:2017nsd,Aaij:2017xva} experiments,
with a lot of properties firstly determined in the recent five years when more than thirty years after the observation of the charmed baryons.
For instance, the absolute branching fractions of two-body non-leptonic charmed baryon decays are measured and collected shown in Table \ref{tab:exp}. Especially, the measurements on the absolute branching fractions of the $\Xi^0_c$ and $\Xi^+_c$ decays by the Belle collaboration \cite{Li:2018qak,Li:2019atu} are important extensions from the studies of $\Lambda_c^+$ decays.
Brilliant prospects of charmed baryon decays are expected at BESIII \cite{BESIII:prospect}, Belle II \cite{Kou:2018nap} and LHCb \cite{Bediaga:2018lhg} in the near future.
Motivated by the experimental progress, many theoretical efforts are devoted to charmed baryon decays since 2016
\cite{Gandhi:2018lez,Zou:2019kzq,Lu:2016ogy,Faustov:2019ddj,Wang:2019dls,Geng:2019awr,Hsiao:2019yur,Geng:2019bfz,Cen:2019ims,
Yang:2019tst,Wang:2017azm,Shi:2017dto,Wang:2018utj,Geng:2018rse,He:2018joe,Grossman:2018ptn,Xie:2016evi,Faustov:2016yza,Li:2016qai,Xie:2017xwx,Geng:2017esc,Geng:2017mxn,Wang:2017gxe,Meinel:2017ggx,Geng:2018plk,
Cheng:2018hwl,Jiang:2018iqa,Zhao:2018zcb,Geng:2018bow,Faustov:2018dkn,Meinel:2016dqj,Geng:2018upx}.
It is worthwhile to investigate the charmed baryon decays continuously since they provide a platform to study the weak and strong interactions.

It is known that the QCD-inspired approaches do not work well in the non-leptonic decays of charmed hadrons due to the fact that the charm quark mass of $1.3$ GeV is neither heavy enough nor light enough.
Except for the model-dependent methods studying charmed baryon decays \cite{Zou:2019kzq,Cheng:2018hwl,Chen:2002jr,Zhao:2018zcb,Cheng:1991sn,Zenczykowski:1993hw,Uppal:1994pt,Fayyazuddin:1996iy},
 the flavor symmetry analysis is model-independent and widely used.
The $SU(3)$ invariant amplitudes are independent of the detailed dynamics and can be determined by fitting experimental data.
Hence the flavor symmetry was usually used soon after some new particles were observed with the decaying dynamics not well-understood, such as the studies on the charm and bottom meson decays in 1970s \cite{Voloshin:1975yx,Gaillard:1974mw,Einhorn:1975fw,Kingsley:1975fe,Wang:1979dx,Eilam:1979mn}, and the singly and doubly charmed baryon decays recently \cite{Gandhi:2018lez,Cen:2019ims,Hsiao:2019yur,Geng:2019awr,Geng:2019bfz,Wang:2019dls,Wang:2018utj,Shi:2017dto,
Wang:2017azm,Geng:2018rse,Grossman:2018ptn,He:2018joe,
Geng:2018upx,Geng:2018bow,Jiang:2018iqa,Geng:2018plk,Wang:2017gxe,Geng:2017mxn,Geng:2017esc,Lu:2016ogy}.

However, the potential of flavor $SU(3)$ symmetry analysis has not been fully explored.
The $SU(3)$ amplitude relations of charmed baryon decays were firstly studied in 1990 \cite{Savage:1989qr} and have to be updated nowadays.
Besides, due to the limited available data and the large number of free parameters in the $SU(3)$ irreducible representation amplitudes, some assumptions are introduced in the global fitting, either by neglecting the $15$-dimensional representation which is small compared to the $6$-dimensional representation, or considering the factorization hypothesis for the $15$-dimensional representation \cite{Cen:2019ims,Geng:2019bfz,Hsiao:2019yur,Geng:2017esc,Geng:2017mxn,Geng:2018upx,Geng:2019awr,Geng:2018rse,Geng:2018bow,Geng:2018plk}. With more and precise experimental data collected in the near future by BESIII, Belle II and LHCb, it deserves to analyze charmed baryon decays based on the exact flavor $SU(3)$ symmetry without any other assumptions to find more and accurate amplitude relations. It is better to test the flavor symmetries and their breaking effects in this way.

The analysis includes the modes charmed baryons decaying into one octet or decuplet light baryon and one pseudoscalar or vector meson, covering almost all the available two-body non-leptonic charm-baryon decays.
Some branching fractions of charmed baryon decays are predicted using the $SU(3)$ amplitude relations.
In order to test the $I$-, $U$-, $V$-spin symmetries and their breaking effects, the amplitude relations are classified according to the $SU(2)$ subgroups of $SU(3)$ group.
We discuss the $\Lambda^+_c\to \Xi^0K^+$, $\Lambda^+_c\to \Sigma^0K^+$, $\Lambda^+_c\to \Sigma^0\pi^+$ modes for testing the $U$-spin symmetry, and the $\Lambda_c^+\to \Sigma^+K^{*0}$ and $\Xi_c^+\to p\overline{K}^{*0}$ modes for the implications of $U$-spin breaking.

\begin{table*}[tp]
\caption{The absolute branching fractions of two-body non-leptonic charmed baryon decays. Data are taken from PDG \cite{Tanabashi:2018oca}, except for some recent results labeled with references.}\label{tab:exp}
\footnotesize\begin{tabular}{|c|c|c||c|c|c|}
\hline \hline
~~~~~~Mode~~~~~~ & ~~~~~Branching fraction~~~~~ & ~~~Type~~~ &   ~~~~~~~~Mode~~~~~~~~ & ~~~~~Branching fraction~~~~~ & ~~~Type~~~\\
\hline
$\Lambda_c^+\to\Lambda^0\pi^+$  & $(1.30\pm0.07)\%$  & CF & $\Lambda_c^+\to\Sigma^{*+}\eta$  & $(1.07\pm0.32)\%$ &CF\\
 $\Lambda_c^+\to\Sigma^0\pi^+$  & $(1.29\pm0.07)\%$  &CF & $\Lambda_c^+\to\Delta^{++}K^-$  & $(1.08\pm0.25)\%$&CF\\
$\Lambda_c^+\to\Sigma^+\pi^0$  & $(1.25\pm0.10)\%$  & CF&$\Lambda_c^+\to\Xi^{*0} K^+$  & $(5.02\pm1.04)\times 10^{-3}$~\cite{Ablikim:2018bir}&CF \\
$\Lambda_c^+\to\Sigma^+\eta$  & $(4.1\pm2.0)\times10^{-3}$~\cite{Ablikim:2018czr} &CF & $\Xi_c^0\to\Xi^-\pi^+$  & $(1.80\pm0.52)\%$~\cite{Li:2018qak} &CF\\
$\Lambda_c^+\to p\overline{K}^0$  & $(3.18\pm0.16)\%$ & CF & $\Lambda_c^+\to\Lambda^0 K^+$ & $(6.1\pm1.2)\times10^{-4}$ & SCS\\
$\Lambda_c^+\to\Xi^0 K^+$  & $(5.5\pm0.7)\times10^{-3}$  & CF&$\Lambda_c^+\to\Sigma^0 K^+$  & $(5.2\pm0.8)\times10^{-4}$ & SCS \\
$\Lambda_c^+\to\Sigma^+\eta^\prime$  &$(1.34\pm0.57)\%$~\cite{Ablikim:2018czr} &CF & $\Lambda_c^+\to p\eta$ & $(1.24\pm0.30)\times10^{-3}$ & SCS \\
$\Lambda_c^+\to\Lambda^0\rho^+$ & $<6\%$ & CF & $\Lambda_c^+\to p \pi^0$ & $<2.7\times10^{-4}$ & SCS \\
 $\Lambda_c^+\to\Sigma^+\rho^0$ & $<1.7\%$ & CF & $\Lambda_c^+\to p\phi$  & $(1.06\pm0.14)\times10^{-3}$&SCS\\
$\Lambda_c^+\to\Sigma^+\phi$ & $(3.9\pm0.6)\times10^{-3}$ & CF &  $\Lambda_c^+\to\Sigma^+ K^{*0}$  & $(3.5\pm1.0)\times10^{-3}$ & SCS\\
$\Lambda_c^+\to p\overline{K}^{*0}$ & $(1.96\pm0.27)\%$ & CF & $\Xi_c^+\to p\overline{K}^{*0}$  & $(2.75\pm1.02)\times 10^{-3}$~\cite{Li:2019atu} & SCS \\
$\Lambda_c^+\to\Sigma^+\omega$ & $(1.70\pm0.21)\%$ &CF &&& \\
\hline\hline
\end{tabular}
\end{table*}

The rest of this paper is organized as follows.
In Sec.~\ref{th}, we introduce the $SU(3)$ irreducible representation amplitude approach and derive the amplitude relations under the $SU(3)_F$ limit.
The Phenomenological analysis is presented in Sec.~\ref{re}.
Sec.~\ref{sum} is a short summary.
The decay amplitudes of charmed baryon decays and a series of amplitude relations under $I$-, $U$-, $V$-spin symmetries are listed in Appendixes \ref{amp} and \ref{relation}, respectively.

\section{Amplitude relations in the flavor symmetry}\label{th}

In this Section, we introduce the $SU(3)$ irreducible representation amplitude (IRA) approach.
The tree level effective
Hamiltonian in charm quark weak decay in the Standard Model (SM) is \cite{Buchalla:1995vs}
 \begin{equation}\label{hsm}
 \mathcal H_{\rm eff}={\frac{G_F}{\sqrt 2} }
\sum_{q=d,s}V_{cq_1}^*V_{uq_2}\big[C_1(\mu)O_1(\mu)+C_2(\mu)O_2(\mu)\big]
,
 \end{equation}
 where $G_F$ is the Fermi coupling constant,
 $C_{1,2}$ are the Wilson coefficients of operators $O_{1,2}$.
$O_{1,2}$ read as
\begin{eqnarray}
O_1=(\bar{u}_{\alpha}q_{2\beta})_{V-A}
(\bar{q}_{1\beta}c_{\alpha})_{V-A},\qquad
O_2=(\bar{u}_{\alpha}q_{2\alpha})_{V-A}
(\bar{q}_{1\beta}c_{\beta})_{V-A},
\end{eqnarray}
in which $\alpha,\beta$ are color indices, $q_{1,2}$ are $d$ and $s$
quarks.
The non-leptonic decays of charmed hadrons are classified into three types: Cabibbo-favored (CF), singly Cabibbo-suppressed (SCS) and doubly Cabibbo-suppressed (DCS) decays,
\begin{equation}
  c\to s\bar d u,\qquad c\to d\bar d /s\bar s u,\qquad c\to d\bar s u.
\end{equation}
 In the $SU(3)$ picture,
the four-quark operators in charm decays embed into an effective Hamiltonian,
\begin{align}\label{h}
  \mathcal{H}_{\rm eff}= \sum_{i,j,k=1}^{3}H^k_{ij}O^{ij}_k,
\end{align}
in which the operator $O^{ij}_k$ is
\begin{align}
  O^{ij}_k = \frac{G_F}{\sqrt 2}\big[C_1(\mu)(\bar{q_i}_{\alpha}q_{k\beta})_{V-A}
(\bar{q}_{j\beta}c_{\alpha})_{V-A}+C_2(\mu)(\bar{q_i}_{\alpha}q_{k\alpha})_{V-A}
(\bar{q}_{j\beta}c_{\beta})_{V-A}\big].
\end{align}
$O^{ij}_k$ can be seen as a tensor representation of $SU(3)$ group.
$H^k_{ij}$ is the corresponding CKM matrix elements of operator $O^{ij}_k$.
Eq.~\eqref{hsm} implies that the tensor
components of $H_{ij}^k$ can be obtained from the map $(\bar uq_1)(\bar q_2c)\rightarrow V^*_{cq_2}V_{uq_1}$ and the others are set to be zero.
 The non-zero components of $H_{ij}^k$ are
\begin{align}\label{ckm1}
 &H_{13}^2 = V_{cs}^*V_{ud},  \qquad H^{2}_{12}=V_{cd}^*V_{ud},\qquad H^{3}_{13}= V_{cs}^*V_{us}, \qquad H^{3}_{12}=V_{cd}^*V_{us}.
\end{align}

Operator $O^{ij}_k$ can be decomposed into four irreducible representations of $SU(3)$ group: $\overline 3 \otimes  3 \otimes \overline3 =  \overline 3\oplus \overline 3^\prime\oplus  6 \oplus \overline{15}$. The explicit decomposition is \cite{Grossman:2012ry,Wang:2019dls}
\begin{align}\label{hd}
  O_k^{ij}= \delta^j_k\Big(\frac{3}{8}O(\overline3)^i-\frac{1}{8}O(\overline3^{\prime})^i\Big)+
  \delta^i_k\Big(\frac{3}{8}O(\overline3^{\prime})^j-\frac{1}{8}O(\overline3)^j\Big)+\epsilon^{ijl}O( 6)_{lk}+O(\overline{15})_k^{ij}.
\end{align}
All components of the irreducible representations can be found in \cite{Wang:2019dls}.
The non-zero components corresponding to tree operators in the $SU(3)$ decomposition, under the approximation of $V_{cs}^*V_{us} = - V_{cd}^*V_{ud}$, are
\begin{align}\label{ckm3}
 &  H( 6)^{22}=-\frac{1}{2}V_{cs}^*V_{ud},\qquad H( 6)^{23}=\frac{1}{4}(V_{cd}^*V_{ud}-V_{cs}^*V_{us}),  \qquad H( 6)^{33}=  \frac{1}{2}V_{cd}^*V_{us},\nonumber \\
 & H(\overline{15})^{2}_{13}= \frac{1}{2}V_{cs}^*V_{ud},  \qquad  H(\overline{15})^{3}_{12}=\frac{1}{2}V_{cd}^*V_{us},\nonumber \\
 &  H(\overline{15})^{2}_{12}= \frac{3}{8}V_{cd}^*V_{ud}-\frac{1}{8}V_{cs}^*V_{us},\qquad H(\overline{15})^{3}_{13}=\frac{3}{8}V_{cs}^*V_{us}-\frac{1}{8}V_{cd}^*V_{ud}.
\end{align}
In the comparison with most literatures \cite{Wang:2017gxe,Cen:2019ims,Geng:2019bfz,Hsiao:2019yur,Geng:2017esc,
Geng:2017mxn,Geng:2018upx,Geng:2019awr,Geng:2018rse,Geng:2018bow,Geng:2018plk,
Wang:2018utj,Wang:2017azm,Savage:1989qr,Shi:2017dto}, the difference is only one common factor of ${1\over2}V_{cs}^*V_{ud}$ or ${1\over4}V_{cs}^*V_{ud}$.

The pseudoscalar and vector mesons form two nonets (octet $+$ singlet),
\begin{align}
P^i_j=
\left(
\begin{array}{ccc}
 \frac{1}{\sqrt6}\eta_8+\frac{1}{\sqrt2}\pi^0 & \pi^+ & K^+\\
 \pi^- & \frac{1}{\sqrt6}\eta_8-\frac{1}{\sqrt2}\pi^0 & K^0\\
 K^- & \overline{K}^0 & -\sqrt{\frac{2}{3}}\eta_8
\end{array}
\right)
+
\frac{1}{\sqrt3}\left(
\begin{array}{ccc}
 \eta_1 & 0 & 0\\
 0 & \eta_1 & 0\\
 0 & 0 & \eta_1
\end{array}
\right),
\end{align}
\begin{align}
V^i_j=
\left(
\begin{array}{ccc}
 \frac{1}{\sqrt6}\omega_8+\frac{1}{\sqrt2}\rho^0 & \rho^+ & K^{*+}\\
 \rho^- & \frac{1}{\sqrt6}\omega_8-\frac{1}{\sqrt2}\rho^0 & K^{*0}\\
 K^{*-} & \overline{K}^{*0} & -\sqrt{\frac{2}{3}}\omega_8
\end{array}
\right)
+
\frac{1}{\sqrt3}\left(
\begin{array}{ccc}
 \omega_1 & 0 & 0\\
 0 & \omega_1 & 0\\
 0 & 0 & \omega_1
\end{array}
\right).
\end{align}
The mass eigenstates $\eta$ and $\eta^\prime$ are mixing of $\eta_8$ and $\eta_1$,
\begin{align}
\left(
\begin{array}{c}
\eta\\
\eta^\prime
\end{array}
\right)
=
\left(
\begin{array}{cc}
\cos\xi  &  -\sin\xi\\
\sin\xi  &  \cos\xi
\end{array}
\right)\left(
\begin{array}{c}
\eta_8\\
\eta_1
\end{array}
\right).
\end{align}
The mixing angle $\xi$ has large uncertainty in literatures. We are not going to discuss the value of $\xi$ because the decay modes involving $\eta$ and $\eta^\prime$ mesons are not used and predicted in this work.
The mass eigenstates $\omega$ and $\phi$ are mixing of $\omega_8$ and $\omega_1$,
\begin{align}
\left(
\begin{array}{c}
\phi\\
\omega
\end{array}
\right)
=
\left(
\begin{array}{cc}
\cos\xi^\prime  &  -\sin\xi^\prime\\
\sin\xi^\prime  &  \cos\xi^\prime
\end{array}
\right)\left(
\begin{array}{c}
\omega_8\\
\omega_1
\end{array}
\right).
\end{align}
The ideal mixing indicates that $\sin\xi^\prime=1/\sqrt3$ and $\cos\xi^\prime=\sqrt{2/3}$.
The singly charmed baryons form an anti-triplet and a sextet. The anti-triplet reads as
\begin{equation}
(\mathcal{B}_{c})_{ij}=
\left(
\begin{array}{ccc}
 0 & \Lambda_c^+ & \Xi_c^+\\
 -\Lambda_c^+ & 0 & \Xi_c^0\\
 -\Xi_c^+ & -\Xi_c^0 & 0
\end{array}
\right),
\end{equation}
or contracted by the Levi-Civita tensor, $(\mathcal{B}_{c})_{ij} = \epsilon_{ijk}(\mathcal{B}_{c})^k$ and $(\mathcal{B}_{c})^k=(\Xi_c^0,-\Xi_c^+,\Lambda_c^+)$.
The light baryon octet reads as
\begin{equation}
(\mathcal{B}_8)^i_j=
\left(
\begin{array}{ccc}
 \frac{1}{\sqrt6}\Lambda^0+\frac{1}{\sqrt2}\Sigma^0 & \Sigma^+ & p\\
 \Sigma^- & \frac{1}{\sqrt6}\Lambda^0-\frac{1}{\sqrt2}\Sigma^0 & n\\
 \Xi^- & \Xi^0 & -\sqrt{\frac{2}{3}}\Lambda^0
\end{array}
\right).
\end{equation}
The light baryon decuplet is symmetric under the interchange of any two quarks, which can be written as
\begin{align}
 &(\mathcal{B}_{10})^{111} = \Delta^{++},  \qquad (\mathcal{B}_{10})^{222}=\Delta^{-},\qquad  (\mathcal{B}_{10})^{333}=\Omega^-, \nonumber\\
 & (\mathcal{B}_{10})^{112} = (\mathcal{B}_{10})^{121} = (\mathcal{B}_{10})^{211} = \frac{1}{\sqrt{3}}\Delta^{+} ,\qquad  (\mathcal{B}_{10})^{122} = (\mathcal{B}_{10})^{212} = (\mathcal{B}_{10})^{221} =  \frac{1}{\sqrt{3}}\Delta^{0},\nonumber\\ & (\mathcal{B}_{10})^{113} =(\mathcal{B}_{10})^{131} =(\mathcal{B}_{10})^{311}=\frac{1}{\sqrt{3}}\Sigma^{*+},\qquad
 (\mathcal{B}_{10})^{223} =(\mathcal{B}_{10})^{232} =(\mathcal{B}_{10})^{322}=\frac{1}{\sqrt{3}}\Sigma^{*-}, \nonumber\\
& (\mathcal{B}_{10})^{133} =(\mathcal{B}_{10})^{313} =(\mathcal{B}_{10})^{331}=\frac{1}{\sqrt{3}}\Xi^{*0},\qquad
 (\mathcal{B}_{10})^{233} =(\mathcal{B}_{10})^{323}=(\mathcal{B}_{10})^{332}=\frac{1}{\sqrt{3}}\Xi^{*-},\nonumber\\
 &(\mathcal{B}_{10})^{123} = (\mathcal{B}_{10})^{132} =(\mathcal{B}_{10})^{213}= (\mathcal{B}_{10})^{231} =(\mathcal{B}_{10})^{312} =(\mathcal{B}_{10})^{321}=\frac{1}{\sqrt{6}}\Sigma^{*0}.
\end{align}
To obtain the $SU(3)$ irreducible representation amplitude of the $\mathcal{B}_c\to \mathcal{B}_8P$ decay, one can contract all indices in the following manner:
\begin{align}\label{amp1}
   \mathcal{A}_{\rm eff}(\mathcal{B}_c\to \mathcal{B}_8P)=&aH(\overline{15})_{jk}^i(\mathcal{B}_{c})^j(\mathcal{B}_8)_l^kP_i^l
   +bH(\overline{15})_{jk}^i(\mathcal{B}_{c})^j(\mathcal{B}_8)_i^lP_l^k
   +cH(\overline{15})_{jk}^i(\mathcal{B}_{c})^l(\mathcal{B}_8)_l^jP_i^k \nonumber\\
   &+dH(\overline{15})_{jk}^i(\mathcal{B}_{c})^l(\mathcal{B}_8)_i^kP_l^j
   +eH(6)^{ij}(\mathcal{B}_{c})_{ik}(\mathcal{B}_8)_l^kP_j^l
   +fH(6)^{ij}(\mathcal{B}_{c})_{ik}(\mathcal{B}_8)_j^lP_l^k\nonumber\\
   &~~+gH(6)^{ij}(\mathcal{B}_{c})_{kl}(\mathcal{B}_8)_i^kP_j^l +hH(\overline{15})^i_{jk}(\mathcal{B}_{c})^j(\mathcal{B}_8)^k_iP^l_l \nonumber\\&~~~~+rH(6)^{ij}(\mathcal{B}_{c})_{ik}(\mathcal{B}_8)^k_jP^l_l.
\end{align}
Similarly, the decay amplitude of $\mathcal{B}_c\to \mathcal{B}_{10}P$ is constructed to be
\begin{align}\label{amp2}
 \mathcal{A}_{\rm eff}(\mathcal{B}_c\to \mathcal{B}_{10}P)=&\alpha (\mathcal{B}_{10})^{ijk}H(\overline{15})^m_{jk}(\mathcal{B}_{c})_{il}P^l_m
   +\beta (\mathcal{B}_{10})^{ijk}H(\overline{15})^l_{jm}(\mathcal{B}_{c})_{il}P^m_k\nonumber\\ &~~
   +\gamma (\mathcal{B}_{10})^{ijk}H(\overline{15})^l_{ij}(\mathcal{B}_{c})_{lm}P^m_k+\delta (\mathcal{B}_{10})^{ijk}H(6)^m_{il}(\mathcal{B}_{c})_{jm}P^l_k\nonumber\\
   &~~~~+\lambda (\mathcal{B}_{10})^{ijk}H(\overline{15})_{jk}^l(\mathcal{B}_{c})_{il}P^m_m.
\end{align}
The decay amplitudes of $a, b, c, e, f, g, h, r$ and $\alpha, \beta, \gamma, \delta, \lambda$ are complex free parameters.
For the decay modes involving vector mesons, their amplitudes have the same forms as Eqs.~\eqref{amp1} and \eqref{amp2}.
For distinguishing, we label superscript $\prime$ for each parameter of vector modes.
With Eqs.~\eqref{amp1} and \eqref{amp2}, the amplitudes of two-body charmed baryon decays are obtained by tensor contraction.
The results are listed in Appendix \ref{amp}.

Another method to analyze the flavor symmetry in charmed baryon decays is direct calculation of the separate $I$-, $U$- and $V$-spin amplitudes. It derives the same amplitude relations as the $SU(3)$ irreducible representation amplitude (IRA) approach. Compared to direct calculation of the $I$-, $U$- and $V$-spin amplitudes, the IRA approach is more operable and programmable. In this work, we use the $SU(3)$ irreducible representation amplitudes to study the symmetry relations between different decay modes. The direct calculation of $I$-, $U$- and $V$-spin amplitudes serves as the verification of our results.
Notice that the octet baryons $\Sigma^0$ and $\Lambda^0$ and the octet mesons $\pi^0$, $\eta_8$, $\rho^0$ and $\omega_8$ do not have definite $U$-spin and $V$-spin quantum numbers. Taking the $\Sigma^0$ and $\Lambda^0$ as examples, they can be written as the mixing of $U$-spin triplet and $U$-spin singlet,
\begin{align}\label{Sigma0}
\Sigma^0=-\frac{1}{2}|1,0\rangle-\frac{\sqrt3}{2}|0,0\rangle,\qquad
  \Lambda^0=\frac{\sqrt3}{2}|1,0\rangle-\frac{1}{2}|0,0\rangle,
\end{align}
or the mixing of $V$-spin triplet and $V$-spin singlet,
\begin{align}
\Sigma^0=\frac{1}{2}|1,0\rangle+\frac{\sqrt3}{2}|0,0\rangle,\qquad
  \Lambda^0=\frac{\sqrt3}{2}|1,0\rangle-\frac{1}{2}|0,0\rangle.
\end{align}

With the decay amplitudes listed in Appendix \ref{amp}, we derive some amplitude relations between different modes.
Here we only list those relations which will be used later. The others are listed in Appendix \ref{relation}.\\
Isospin relations:
\begin{equation}\label{Isospin1}
\mathcal{A}(\Lambda_c^+\rightarrow\Sigma^+\pi^0)+\mathcal{A}(\Lambda_c^+\rightarrow\Sigma^0\pi^+)=0,
\end{equation}
\begin{equation}\label{Isospin2}
\mathcal{A}(\Lambda_c^+\rightarrow\Sigma^+\rho^0)+\mathcal{A}(\Lambda_c^+\rightarrow\Sigma^0\rho^+)=0,
\end{equation}
\begin{equation}\label{Isospin3}
\mathcal{A}(\Lambda_c^+\rightarrow\Delta^{++}K^-)-\sqrt{3}\mathcal{A}(\Lambda_c^+\rightarrow\Delta^+\overline{K}^0)=0.
\end{equation}
$U$-spin relations:
\begin{equation}\label{Uspin1}
\sqrt2\sin\theta\mathcal{A}(\Lambda^+_c\rightarrow\Sigma^0\pi^+) -\sqrt2\mathcal{A}(\Lambda^+_c\rightarrow\Sigma^0K^+)
+\sin\theta\mathcal{A}(\Lambda^+_c\rightarrow\Xi^0K^+)= 0,
\end{equation}
\begin{equation}\label{Uspin9}
\mathcal{A}(\Lambda_c^+\rightarrow \Sigma^+K^{*0})
-\mathcal{A}(\Xi_c^+\rightarrow p\overline{K}^{*0})=0,
\end{equation}
\begin{equation}\label{Uspin2}
\sin\theta\mathcal{A}(\Lambda_c^+\rightarrow p\overline{K}^0)
-\sqrt2\mathcal{A}(\Lambda_c^+\rightarrow p\pi^0)
+\sqrt2\sin\theta\mathcal{A}(\Lambda_c^+\rightarrow\Sigma^+\pi^0)=0,
\end{equation}
\begin{equation}\label{Uspin3}
\sin^2\theta\mathcal{A}(\Lambda_c^+\rightarrow\Xi^0K^+)
-\mathcal{A}(\Xi_c^+\rightarrow n\pi^+)=0,
\end{equation}
\begin{equation}\label{Uspin4}
\sin^2\theta\mathcal{A}(\Lambda_c^+\rightarrow p\overline{K}^0)
-\mathcal{A}(\Xi_c^+\rightarrow\Sigma^+K^0)=0,
\end{equation}
\begin{equation}\label{Uspin5}
\sin^2\theta\mathcal{A}(\Lambda_c^+\rightarrow p\overline{K}^{*0})
+\mathcal{A}(\Lambda_c^+\rightarrow pK^{*0})
-\sin\theta\mathcal{A}(\Lambda_c^+\rightarrow\Sigma^+K^{*0})=0,
\end{equation}
\begin{equation}\label{Uspin6}
\sin\theta\mathcal{A}(\Lambda_c^+\rightarrow p\overline{K}^{*0})
-\sqrt2\mathcal{A}(\Lambda_c^+\rightarrow p\rho^0)
+\sqrt2\sin\theta\mathcal{A}(\Lambda_c^+\rightarrow\Sigma^+\rho^0)=0,
\end{equation}
\begin{equation}\label{Uspin7}
\sin^2\theta\mathcal{A}(\Lambda_c^+\rightarrow p\overline{K}^{*0})
-\mathcal{A}(\Xi_c^+\rightarrow\Sigma^+K^{*0})=0,
\end{equation}
\begin{equation}\label{Uspin8}
\mathcal{A}(\Lambda_c^+\rightarrow pK^{*0})
-\sin^2\theta\mathcal{A}(\Xi_c^+\rightarrow\Sigma^+\overline{K}^{*0})=0,
\end{equation}
\begin{equation}\label{Uspin10}
\sqrt2\sin^2\theta\mathcal{A}(\Lambda_c^+\rightarrow\Sigma^+\rho^0)
+\sin\theta\mathcal{A}(\Lambda_c^+\rightarrow\Sigma^+K^{*0})
-\sqrt2\mathcal{A}(\Xi_c^+\rightarrow p\rho^0)=0,
\end{equation}
\begin{equation}\label{Uspin11}
\sin^2\theta\mathcal{A}(\Lambda_c^+\rightarrow\Delta^+\overline{K}^0)
+\mathcal{A}(\Xi_c^+\rightarrow\Sigma^{*+}K^0)=0,
\end{equation}
\begin{align}\label{Uspin12}
&\sin^2\theta\mathcal{A}(\Lambda_c^+\rightarrow\Delta^{++}K^-)
=\sin\theta\mathcal{A}(\Lambda_c^+\rightarrow\Delta^{++}\pi^-)\nonumber\\
&~~~~~~~~~~~~~~~
=-\sin\theta\mathcal{A}(\Xi_c^+\rightarrow\Delta^{++}K^-)=-\mathcal{A}(\Xi_c^+\rightarrow\Delta^{++}\pi^-),
\end{align}
\begin{align}\label{Uspin13}
&\sin^2\theta\mathcal{A}(\Xi_c^0 \rightarrow \Xi^- \pi^+)
=-\sin\theta\mathcal{A}(\Xi_c^0 \rightarrow \Sigma^-\pi^+)\nonumber\\
&~~~~~~~~~~~~~~~=\sin\theta\mathcal{A}(\Xi_c^0 \rightarrow \Xi^- K^+)=-\mathcal{A}(\Xi_c^0 \rightarrow \Sigma^- K^+),
\end{align}
\begin{equation}\label{Uspin14}
\sin^2\theta\mathcal{A}(\Lambda_c^+\rightarrow\Xi^{*0}K^+)
+\mathcal{A}(\Xi_c^+\rightarrow\Delta^0\pi^+)=0,
\end{equation}
\begin{equation}\label{aXic+1}
\mathcal{A}(\Lambda_c^+\to nK^+)-\sin^2\theta\mathcal{A}(\Xi_c^+\to\Xi^0\pi^+)=0,
\end{equation}
\begin{equation}\label{aXic+2}
\mathcal{A}(\Lambda_c^+\to\Delta^+K^0)+\sin^2\theta\mathcal{A}(\Xi_c^+\to\Sigma^{*+}\overline{K}^0)=0,
\end{equation}
\begin{equation}\label{aXic+3}
\mathcal{A}(\Lambda_c^+\to\Delta^0K^+)+\sin^2\theta\mathcal{A}(\Xi_c^+\to\Xi^{*0}\pi^+)=0.
\end{equation}
$V$-spin relation:
\begin{equation}\label{Vspin}
 \mathcal{A}(\Xi_c^+\to\Sigma^{*+}\overline{K}^0)+\mathcal{A}(\Xi_c^+\to\Xi^{*0}\pi^+)=0.
\end{equation}
In above equations, $\theta$ is the Cabibbo angle and $\sin\theta \simeq V_{us}$.

One can derive more amplitude relations that are only valid in the flavor $SU(3)$ symmetry (but no longer valid in one of the $SU(2)$ subgroups) by combining two or three relations belonging to different $SU(2)$ subgroups.
In principle, if all the amplitude relations belonging to the $I$-, $U$-, $V$-spins are found, all the amplitude relations in the flavor $SU(3)$ symmetry can be obtained by combining those amplitude relations belonging to the three $SU(2)$ subgroups because the generators of three $SU(2)$ subgroups contain all the  generators of $SU(3)$ group.

In the $U$-spin relations, one type of them, which is relevant to a complete interchange of $d$ and $s$ quarks in the initial and final states in two decay channels, is simplest. For example, under the complete interchange of $d\leftrightarrow s$, the initial and final state particles in Eq.~\eqref{Uspin9} are interchanged as
\begin{align}
  \Lambda^+_c \leftrightarrow \Xi^+_c, \qquad p \leftrightarrow \Sigma^+,\qquad K^{*0} \leftrightarrow \overline K^{*0}.
\end{align}
The relations associated with the complete interchange of $d$ and $s$ quarks are very interesting because they can be gotten from their initial and final states without writing down the amplitude decompositions and the $CP$ asymmetries in the two decay modes are summed to be zero in the $U$-spin limit \cite{Wang:2019dls}.

All the symmetry relations listed in Eqs.~\eqref{Isospin1} $\sim$ \eqref{aXic+3} and Appendix \ref{relation} can be examined by the direct calculation of the $I$-, $U$- and $V$-spin amplitudes. Taking  Eq.~\eqref{Uspin1} as an example, we show the computational procedure in detail.
The final-state mesons and baryons $\pi^+$, $K^+$, $\Xi^0$ and $\Sigma^0$ can be written as $U$-spin multiplets: $|\pi^+\rangle=-|\frac{1}{2},-\frac{1}{2}\rangle$, $|K^+\rangle=|\frac{1}{2},\frac{1}{2}\rangle$, $|\Xi^0\rangle=|1,-1\rangle$ and $|\Sigma^0\rangle=-\frac{1}{2}|1,0\rangle-\frac{\sqrt3}{2}|0,0\rangle$.
The effective Hamiltonian of CF decay changes the $U$-spin and its third component, $|\mathcal{H}^{\rm CF}_{\rm eff}\rangle=-|1,-1\rangle$. Then we have
\begin{equation}
\mathcal{H}_{\rm eff}^{\rm CF}|\Lambda_c^+\rangle=
-|1,-1;\frac{1}{2},\frac{1}{2}\rangle=
-\frac{1}{\sqrt3}|\frac{3}{2},-\frac{1}{2}\rangle+\sqrt{\frac{2}{3}}|\frac{1}{2},-\frac{1}{2}\rangle.
\end{equation}
The $U$-spin representations of $|\Sigma^0\pi^+\rangle$ and $|\Xi^0K^+\rangle$ states are
\begin{equation}
|\Sigma^0\pi^+\rangle=\frac{1}{2}|1,0;\frac{1}{2},-\frac{1}{2}\rangle+\frac{\sqrt3}{2}|0,0;\frac{1}{2},-\frac{1}{2}\rangle
=\frac{1}{\sqrt6}|\frac{3}{2},-\frac{1}{2}\rangle+\frac{1}{2\sqrt3}|\frac{1}{2},-\frac{1}{2}\rangle^{(1)}+\frac{\sqrt3}{2}|\frac{1}{2},-\frac{1}{2}\rangle^{(2)},
\end{equation}
\begin{equation}
|\Xi^0K^+\rangle=|1,-1;\frac{1}{2},\frac{1}{2}\rangle=
\frac{1}{\sqrt3}|\frac{3}{2},-\frac{1}{2}\rangle-\sqrt{\frac{2}{3}}|\frac{1}{2},-\frac{1}{2}\rangle^{(1)}.
\end{equation}
Then the decay amplitudes of $\Lambda_c^+\to\Sigma^0\pi^+$ and  $\Lambda_c^+\to\Xi^0K^+$ can be expressed as
\begin{equation}\label{p1}
\mathcal{A}(\Lambda_c^+\to\Sigma^0\pi^+)=\langle\Sigma^0\pi^+|\mathcal{H}^{\rm CF}_{\rm eff}|\Lambda_c^+\rangle=-\frac{\sqrt2}{6}\mathcal{A}_{\frac{3}{2}}+
\frac{\sqrt2}{6}\mathcal{A}_{\frac{1}{2}}^{(1)}+\frac{1}{\sqrt2}\mathcal{A}_{\frac{1}{2}}^{(2)},
\end{equation}
\begin{equation}\label{p2}
\mathcal{A}(\Lambda_c^+\to\Xi^0K^+)=\langle\Xi^0K^+|\mathcal{H}^{\rm CF}_{\rm eff}|\Lambda_c^+\rangle=-\frac{1}{3}\mathcal{A}_{\frac{3}{2}}-\frac{2}{3}\mathcal{A}_{\frac{1}{2}}^{(1)}.
\end{equation}
The effective Hamiltonian of SCS decay $\Lambda_c^+\to\Sigma^0K^+$ can be written as $|\mathcal{H}_{\rm eff}^{\rm SCS}\rangle=\sqrt2|1,0\rangle$ since the SCS transition is $c\to(s\bar s-d\bar d)u $ and the minus sign between $s\bar s$ and $d\bar d$ arises from the approximation of the CKM matrix elements $V_{cd}^*V_{ud}=-V_{cs}^*V_{us}$. And then
\begin{equation}
\mathcal{H}_{\rm eff}^{\rm SCS}|\Lambda_c^+\rangle=
\sqrt{2}|1,0;\frac{1}{2},\frac{1}{2}\rangle=
\frac{2}{\sqrt3}|\frac{3}{2},\frac{1}{2}\rangle-\sqrt{\frac{2}{3}}|\frac{1}{2},\frac{1}{2}\rangle.
\end{equation}
The final state $|\Sigma^0K^+\rangle$ can be written as $U$-spin multiplets as
\begin{equation}
|\Sigma^0K^+\rangle=-\frac{1}{2}|1,0;\frac{1}{2},\frac{1}{2}\rangle-\frac{\sqrt3}{2}|0,0;\frac{1}{2},\frac{1}{2}\rangle
=-\frac{1}{\sqrt6}|\frac{3}{2},\frac{1}{2}\rangle+\frac{1}{2\sqrt3}|\frac{1}{2},\frac{1}{2}\rangle^{(1)}-\frac{\sqrt3}{2}|\frac{1}{2},\frac{1}{2}\rangle^{(2)}.
\end{equation}
The decay amplitude of $\Lambda_c^+\to\Sigma^0K^+$ decay reads as
\begin{equation}\label{p3}
\mathcal{A}(\Lambda_c^+\to\Sigma^0K^+)=\langle\Sigma^0K^+|\mathcal{H}_{\rm eff}^{\rm SCS}|\Lambda_c^+\rangle=-\frac{\sqrt2}{3}\mathcal{A}_{\frac{3}{2}}-\frac{\sqrt2}{6}\mathcal{A}_{\frac{1}{2}}^{(1)}+\frac{1}{\sqrt2}\mathcal{A}_{\frac{1}{2}}^{(2)}.
\end{equation}
According to Eqs.~\eqref{p1}, \eqref{p2} and \eqref{p3}, the $U$-spin relation Eq.~\eqref{Uspin1} is confirmed.
Another example is the $V$-spin symmetry relation~\eqref{Vspin}.
The initial state $\Xi_c^+$ is a $V$-spin singlet. The final-state mesons $\pi^+$ and $\overline{K}^0$ form a $V$-spin doublet $(\pi^+,\overline{K}^0)$ and baryons $\Sigma^{*+}$ and $\Xi^{*0}$ are included in the $V$-spin multiplets $(\Delta^{++},\Sigma^{*+},\Xi^{*0},\Omega^-)$.
The Cabibbo-favored effective Hamiltonian changes the $V$-spin by  $|\frac{1}{2},-\frac{1}{2}\rangle|\frac{1}{2},\frac{1}{2}\rangle=\frac{1}{\sqrt2}|1,0\rangle-\frac{1}{\sqrt2}|0,0\rangle$.
We can derive
\begin{equation}
 \mathcal{H}_{\rm eff}^{\rm CF}|\Xi_c^+\rangle=(\frac{1}{\sqrt2}|1,0\rangle-\frac{1}{\sqrt2}|0,0\rangle)|0,0\rangle
 =\frac{1}{\sqrt2}|1,0\rangle-\frac{1}{\sqrt2}|0,0\rangle,
\end{equation}
\begin{equation}
 |\Sigma^{*+}\overline{K}^0\rangle=|\frac{3}{2},\frac{1}{2}\rangle|\frac{1}{2},-\frac{1}{2}\rangle
 =\frac{1}{\sqrt2}|2,0\rangle-\frac{1}{\sqrt2}|1,0\rangle,
\end{equation}
\begin{equation}
 |\Xi^{*0}\pi^+\rangle=|\frac{3}{2},-\frac{1}{2}\rangle|\frac{1}{2},\frac{1}{2}\rangle
 =\frac{1}{\sqrt2}|2,0\rangle+\frac{1}{\sqrt2}|1,0\rangle.
\end{equation}
Then the decay amplitudes of $\Xi_c^+\to\Sigma^{*+}\overline{K}^0$ and $\Xi_c^+\to\Xi^{*0}\pi^+$ are
\begin{equation}
 \mathcal{A}(\Xi_c^+\to\Sigma^{*+}\overline{K}^0)=\langle\Sigma^{*+}\overline{K}^0|\mathcal{H}_{\rm eff}^{\rm CF}|\Xi_c^+\rangle=-\frac{1}{2}\mathcal{A}_1,
\end{equation}
\begin{equation}
 \mathcal{A}(\Xi_c^+\to\Xi^{*0}\pi^+)=\langle\Xi^{*0}\pi^+|\mathcal{H}^{\rm CF}_{\rm eff}|\Xi_c^+\rangle=\frac{1}{2}\mathcal{A}_1,
\end{equation}
being consistent with Eq.~\eqref{Vspin}

\section{Phenomenological analysis}\label{re}

\subsection{Test flavor symmetry}

In this Section, we discuss physical applications of the amplitude relations in the $SU(3)_F$ limit.
For the two-body decay, for instance $\mathcal{B}_c\to \mathcal{B}_8P$, the partial decay width $\Gamma$ is  parameterized to be \cite{Pakvasa:1990if}
\begin{equation}\label{width}
\Gamma(\mathcal{B}_c\to \mathcal{B}_8P)=\frac{\left| P_c\right|}{8\pi m_{\mathcal{B}_c}^2}\big\{[(m_{\mathcal{B}_c}+m_{\mathcal{B}_8})^2-m_P^2]\left|S\right|^2+
[(m_{\mathcal{B}_c}-m_{\mathcal{B}_8})^2-m_P^2]\left|P\right|^2\big\},
\end{equation}
where $S/P$ is the $S/P$-wave amplitude and $\left| P_c\right|$ is the $c.m.$ momentum in the rest frame of initial state,
\begin{align}
\left| P_c\right| =\frac{\sqrt{[m_{\mathcal{B}_c}^2-(m_{\mathcal{B}_8}+m_P)^2]
[m_{\mathcal{B}_c}^2-(m_{\mathcal{B}_8}-m_P)^2]}}{2m_{\mathcal{B}_c}}.
\end{align}
To test the $SU(3)_F$ relations via branching fractions, the parameters before $\left|S\right|^2$ and $\left|P\right|^2$ in Eq.~\eqref{width} are assumed to follow the flavor symmetry and hence can be absorbed into the decay amplitudes,
\begin{equation}\label{gam}
\Gamma(\mathcal{B}_c\to \mathcal{B}_8P)\simeq\frac{\left|P_c\right|}{8\pi m_{\mathcal{B}_c}^2}\big(\left|S^\prime\right|^{2}+\left|P^\prime\right|^{2}\big)=\frac{\left|P_c\right|}{8\pi m_{\mathcal{B}_c}^2}\left|\mathcal{A}\right|^2.
\end{equation}
The partial decay width of $\mathcal{B}_c\to \mathcal{B}_8P$ can be parameterized to be \cite{Pakvasa:1990if}
\begin{align}
 \Gamma(\mathcal{B}_c\to \mathcal{B}_8V)&=\frac{\left| P_c\right|}{4\pi}\frac{E_{\mathcal{B}_8}+m_{\mathcal{B}_8}}
 {m_{\mathcal{B}_c}}[2(\left|S\right|^2+\left|P_2\right|^2)+\frac{E_V^2}{m_V^2}
 (\left|S+D\right|^2+\left|P_1\right|^2)].
\end{align}
Similar to $\mathcal{B}_c\to \mathcal{B}_8P$, we can write
\begin{align}
 \Gamma(\mathcal{B}_c\to \mathcal{B}_8V)
 \simeq\frac{\left|P_c\right|}{8\pi m_{\mathcal{B}_c}^2}\left|\mathcal{A}\right|^2,
\end{align}
\begin{align}
\left|\mathcal{A}\right|^2=
m_{\mathcal{B}_c}(E_{\mathcal{B}_8}+m_{\mathcal{B}_8})
[4(\left|S\right|^2+\left|P_2\right|^2)+2\frac{E_V^2}{m_V^2}
(\left|S+D\right|^2+\left|P_1\right|^2)],
\end{align}
and assume amplitude $\mathcal{A}$ follows the flavor symmetry.
The similar trick is also used in the modes involving baryon decuplet for simplification.

The first application of the amplitude relations is to test the flavor symmetry.
According to Eq.~\eqref{gam} and the Isospin symmetry relation~\eqref{Isospin1}, the ratio of $\mathcal{B}r(\Lambda_c^+\to\Sigma^+\pi^0)$ and $\mathcal{B}r(\Lambda_c^+\to\Sigma^0\pi^+)$ is calculated to be
\begin{equation}
  \mathcal{B}r(\Lambda_c^+\to\Sigma^+\pi^0)/\mathcal{B}r(\Lambda_c^+\to\Sigma^0\pi^+)=1.00.
\end{equation}
The experimental data of  $\mathcal{B}r(\Lambda_c^+\to\Sigma^+\pi^0)$ and $\mathcal{B}r(\Lambda_c^+\to\Sigma^0\pi^+)$ imply that \cite{Tanabashi:2018oca}
\begin{equation}\label{ib}
  \mathcal{B}r(\Lambda_c^+\to\Sigma^+\pi^0)/\mathcal{B}r(\Lambda_c^+\to\Sigma^0\pi^+)=0.96\pm0.09.
\end{equation}
One can find the theoretical prediction is well consistent with the experimental data.
It demonstrates that the isospin symmetry is fairly accurate even in the charmed baryon decays.

\begin{figure}
  \centering
  \includegraphics[width=0.35\textwidth]{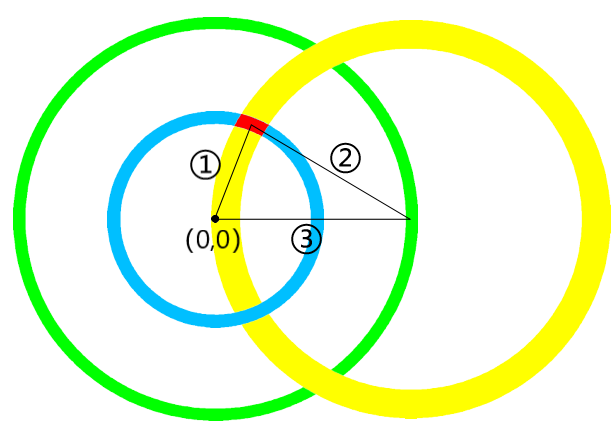}\\
\caption{Triangle constructed by the amplitudes of \textcircled
{1}\,\,$\Lambda_c^+\to\Xi^0K^+$, \textcircled
{2}\,\,$\Lambda_c^+\to\Sigma^0K^+$ and \textcircled
{3}\,\,$\Lambda_c^+\to\Sigma^0\pi^+$, in which the circular rings present the errors of the three amplitudes. }
\label{tx}
\end{figure}

Other testable equation is the $U$-spin relation \eqref{Uspin1}. The amplitude magnitudes of $\Lambda_c^+\to\Sigma^0\pi^+$, $\Lambda_c^+\to\Sigma^0K^+$ and $\Lambda_c^+\to\Xi^0K^+$ modes obtained from available data are
\begin{align}
\sin\theta|\mathcal{A}(\Lambda_c^+\to\Xi^0K^+)|&=(4.29\pm0.28)\times10^{-7}{\rm GeV},\nonumber\\
  \sqrt2|\mathcal{A}(\Lambda_c^+\to\Sigma^0K^+)|&=(7.82\pm0.61)\times10^{-7}{\rm GeV},\nonumber\\
  \sqrt2\sin\theta|\mathcal{A}(\Lambda_c^+\to\Sigma^0\pi^+)|&=(8.28\pm0.26)\times10^{-7}{\rm GeV}.
\end{align}
If the $U$-spin symmetry is relatively precise, amplitudes of the three modes should form a triangle in the complex plane. The triangle is shown in Fig.~\ref{tx}.
It is found the amplitudes of $\Lambda_c^+\to\Sigma^0\pi^+$, $\Lambda_c^+\to\Sigma^0K^+$ and $\Lambda_c^+\to\Xi^0K^+$ modes form a triangle within the $1\sigma$ error.
Using the triangle in Fig.~\ref{tx}, we extract the relative strong phases between the three decay modes from data. The three angles of the triangle are expected to be
\begin{align}
  \theta(\textcircled{1})=30.74^\circ\pm2.32^\circ,\qquad
  \theta(\textcircled{2})=68.70^\circ\pm8.82^\circ,\qquad
  \theta(\textcircled{3})=80.56^\circ\pm8.52^\circ,
\end{align}
in which $\theta(\textcircled{1},\,\textcircled{2},\,\textcircled{3} )$ present the opposite angles of the three sides of the triangle in Fig.~\ref{tx}. These values could provide some guides for the theoretical studies.

\subsection{Predictions for branching fractions}
From last subsection, one can find the flavor $SU(3)$ symmetry is a reliable tool to study charmed baryon decays.
With the amplitude relations given in Eqs.~\eqref{Isospin1} $\sim$ \eqref{aXic+3}, we estimate some branching fractions of charmed baryon decays. Our results are presented in Table \ref{tab}.  For the branching fractions extracted from the symmetry relations with three decay modes, such as  $\mathcal{B}r(\Xi_c^+\to\Sigma^+\overline K^{*0})$, the upper and lower limit are obtained via the property of triangle that the sum of the two sides is greater than the third side and the difference between the two sides is less than the third side.
In Refs.~\cite{Cen:2019ims,Geng:2019bfz,Hsiao:2019yur,Geng:2017esc,Geng:2017mxn,Geng:2018upx,Geng:2019awr,Geng:2018rse,Geng:2018bow,Geng:2018plk}, the contributions from $O(\overline {15})$ being negligible compared to $O(6)$ is used in their predictions. This approximation is questionable because the ratio of the Wilson coefficients of $O(6)$ and $O(\overline {15})$, $C_-/C_+ \approx 2.4$ \cite{Gaillard:1974nj,Altarelli:1974exa}, is not large enough. For comparison, we list the results given in \cite{Hsiao:2019yur,Geng:2019awr,Geng:2018rse,Geng:2018plk}
in Table \ref{tab}. From Table \ref{tab}, one can find our results are consistent with Refs.~\cite{Hsiao:2019yur,Geng:2019awr,Geng:2018rse,Geng:2018plk} in most decay modes.
In Table \ref{tab}, the branching fraction of $\Lambda_c^+\to\Delta^+\overline{K}^0$ is the most precise prediction because it is derived from Isospin symmetry.
Our prediction of
$\mathcal{B}r(\Lambda_c^+\to p\pi^0)$ is given in a large range which satisfies the upper limit by BESIII Collaboration \cite{Ablikim:2017ors}, $\mathcal{B}r(\Lambda^+_c\to p\pi^0)<2.7\times 10^{-4}$ in $90\%$ confidence level.

\begin{table}[t!]
\caption{Predictions for branching fractions of charmed baryon decays. Our results are compared to the results given in \cite{Hsiao:2019yur,Geng:2019awr,Geng:2018rse,Geng:2018plk}
in which the approximation that $O(\overline {15})$ contributions are negligible compared to $O(6)$ is used.}\label{tab}
\centering
\begin{tabular}{|c|c|c|c|}
\hline\hline
Mode & This work & ~~~~Type~~~~  & Geng $et~al.$  \\
\hline
 $\Lambda_c^+\to\Sigma^0\rho^+$   &  $<(1.69\pm0.07)\%$  & CF & $(6.1\pm4.6)\times10^{-3}$~\cite{Hsiao:2019yur}   \\
$\Lambda_c^+\to\Delta^+\overline K^0$  & $(3.59\pm0.85)\times10^{-3}$   & CF & $(5.1\pm0.8)\times10^{-3}$~\cite{Geng:2019awr} \\
$\Xi_c^+\to\Sigma^+\overline K^{*0}$  &  $(5.75\pm2.18)\%<\mathcal{B}r<(38.19\pm6.16)\%$   & CF & $(10.1\pm2.9)\%$~\cite{Hsiao:2019yur} \\
$\Lambda_c^+\to p\pi^0$   &  ~$(6.90\pm7.56)\times10^{-6}<\mathcal{B}r<(3.18\pm0.19)\times10^{-3}$~   & SCS & $(1.3\pm0.7)\times10^{-4}$~\cite{Geng:2018rse} \\
$\Lambda_c^+\to p\rho^0$  &  $(1.03\pm0.35)\times10^{-4}<\mathcal{B}r<(3.25\pm0.22)\times10^{-3}$   & SCS &  $(3.5\pm2.9)\times10^{-4}$~\cite{Hsiao:2019yur}  \\
$\Lambda_c^+\to\Delta^{++}\pi^-$  &  $(6.19\pm1.46)\times10^{-4}$   & SCS & $(12.5\pm2.0)\times10^{-4}$~\cite{Geng:2019awr}  \\
$\Xi_c^0\to\Sigma^-\pi^+$ & $(9.78\pm3.25)\times10^{-4}$ &SCS & $(9.0\pm0.4)\times10^{-4}$~\cite{Geng:2018plk}  \\
$\Xi_c^0\to\Xi^-K^+$ & $(8.23\pm2.73)\times10^{-4}$ & SCS& $(7.6\pm0.4)\times10^{-4}$~\cite{Geng:2018plk}  \\
~$\Xi_c^+\to\Delta^{++}K^-$~  &   $(1.23\pm0.30)\times10^{-3}$  & SCS  &   $(3.5\pm0.6)\times10^{-3}$~\cite{Geng:2019awr}  \\
$\Lambda_c^+\to p K^{*0}$  &   $(8.09\pm4.26)\times10^{-5}<\mathcal{B}r<(5.37\pm1.11)\times10^{-4}$   & DCS &$(1.6\pm0.5)\times10^{-4}$~\cite{Hsiao:2019yur} \\
$\Xi_c^0\to\Sigma^-K^+$ & $(4.55\pm1.51)\times10^{-5}$ & DCS& $(4.5\pm0.2)\times10^{-5}$~\cite{Geng:2018plk}  \\
$\Xi_c^+\to n\pi^+$  &  ~~$(4.30\pm0.62)\times10^{-5}$~~ & DCS  &  ~$(4.76\pm1.22)\times10^{-5}$~\cite{Geng:2018plk} \\
$\Xi_c^+\to\Sigma^+ K^0$  &  $(1.53\pm0.13)\times10^{-4}$    & DCS & $(1.76\pm0.08)\times10^{-4}$~\cite{Geng:2018plk}   \\
$\Xi_c^+\to p\rho^0$  &  $(4.33\pm3.39)\times10^{-5}<\mathcal{B}r<(8.47\pm1.58)\times10^{-4}$  & DCS &$(7.1\pm2.2)\times10^{-5}$~\cite{Hsiao:2019yur}  \\
$\Xi_c^+\to\Sigma^+K^{*0}$   &  $(9.16\pm1.40)\times10^{-5}$   & DCS & $(9.9\pm1.3)\times10^{-5}$~\cite{Hsiao:2019yur}  \\
$\Xi_c^+\to\Sigma^{*+} K^0$  &  $(1.84\pm0.45)\times10^{-5}$  & DCS  &   $(3.5\pm0.6)\times10^{-5}$~\cite{Geng:2019awr}   \\
$\Xi_c^+\to\Delta^{++}\pi^-$   &  $(6.80\pm1.64)\times10^{-5}$   &DCS & $(25.5\pm4.4)\times10^{-5}$~\cite{Geng:2019awr} \\
$\Xi_c^+\to\Delta^0\pi^+$  &  $(4.73\pm1.03)\times10^{-5}$  &DCS &  $(8.5\pm1.5)\times10^{-5}$~\cite{Geng:2019awr}  \\
\hline\hline
\end{tabular}
\end{table}

\begin{table}[t!]
\caption{Branching fractions of $\Xi_c^+$ decays predicted via the relative ratios to $\Xi_c^+\to \Xi^-\pi^+\pi^+$.}\label{tabx}
\centering
\footnotesize\begin{tabular}{|c|c|c|c|c|}
\hline\hline
Modes & ~Ratio (relative to $\Xi_c^+\to\Xi^-\pi^+\pi^+$)~  & ~Branching
fraction~ & ~~~~Type~~~~ &  Geng $et~al.$\\
\hline
$\Xi_c^+\to\Sigma^{*+}\overline K^0$  & $1.0\pm0.5$   & $(2.86\pm1.91)\%$  & CF  & ---\\
$\Xi_c^+\to \Sigma^+\overline{K}^{*0}$   &  $0.81\pm0.15$  & $(2.32\pm1.11)\%$ & CF & $(10.1\pm2.9)\%$~\cite{Hsiao:2019yur}\\
$\Xi_c^+\to\Xi^0\pi^+$  &  $0.55\pm0.16$   & $(1.57\pm0.84)\%$   & CF  & $(8.1\pm4.0)\times10^{-3}$~\cite{Geng:2018plk}\\
$\Xi_c^+\to\Xi^{*0}\pi^+$  &  $<0.10$   &  ~$<(2.86\pm1.27)\times10^{-3}$~   & CF & ---\\
$\Xi_c^+\to\Sigma^+\phi$  &  $<0.11$    &   $<(3.15\pm1.40)\times10^{-3}$  & SCS  & ~$(1.9\pm0.9)\times10^{-3}$~\cite{Hsiao:2019yur}\\
\hline\hline
\end{tabular}
\end{table}

\begin{table}[t!]
\caption{Branching fractions predicted via  Table~\ref{tabx}.}\label{taby}
\centering
\footnotesize\begin{tabular}{|c|c|c|}
\hline\hline
~~Modes~~ & ~~~~~~~~~~~Branching fraction~~~~~~~~~~~~ & ~~~~Type~~~~ \\
\hline
$\Lambda_c^+\to nK^+$   &   $(2.11\pm1.14)\times10^{-5}$ & DCS \\
$\Lambda_c^+\to\Delta^+K^0$  & $(3.67\pm2.46)\times10^{-5}$  & DCS\\
$\Lambda_c^+\to\Delta^0K^+$  & $<(3.65\pm1.64)\times10^{-6}$ & DCS\\
\hline\hline
\end{tabular}
\end{table}

There are some branching fraction ratios relative to $\mathcal{B}r(\Xi_c^+\to\Xi^-\pi^+\pi^+)$ given by PDG~\cite{Tanabashi:2018oca}.
For example, the ratio between $\mathcal{B}r(\Xi_c^+\to \Sigma^{*+}\overline{K}^0)$ and $\mathcal{B}r(\Xi_c^+\to\Xi^-\pi^+\pi^+)$ is
\begin{align}
\mathcal{B}r(\Xi_c^+\to \Sigma^{*+}\overline{K}^0)/\mathcal{B}r(\Xi_c^+\to\Xi^-\pi^+\pi^+) = 1.0\pm0.5.
\end{align}
The branching fraction of $\Xi_c^+\to\Xi^-\pi^+\pi^+$ is taken from \cite{Li:2019atu}.
And then the branching fractions, such as $\mathcal{B}r(\Xi_c^+\to \Sigma^{*+}\overline{K}^0)$, can be predicted using these ratios.
The results are presented in Table~\ref{tabx}.
With the results listed in Table.~\ref{tabx}, one can also predict some branching fractions via amplitude relations.
The results are presented in Table~\ref{taby}.
From Table~\ref{tab} and Table~\ref{tabx}, one can find the predictions of $\mathcal{B}r(\Xi_c^+\to\Sigma^+\overline{K}^{*0})$ in two different methods are consistent with each other within the range of errors and biased to the smaller value.

\subsection{$K^0_S-K^0_L$ asymmetry and $CP$ asymmetry in $\mathcal{B}_c\to \mathcal{B}K^0_{S,L}$ decays}

Flavor $SU(3)$ symmetry can give some interesting arguments for the $K^0_S-K^0_L$ asymmetry and $CP$ asymmetry in charm hadron decays into neutral kaons. For convenience to the analysis below, we first review the key points about $K^0_S-K^0_L$ asymmetry and $CP$ asymmetry in charm hadron decays into neutral kaons. More details can be found in \cite{Yu:2017oky,Wang:2017gxe,Wang:2017ksn}.
The $K_S^0-K_L^0$ asymmetry, which is induced by the interference between CF and DCS amplitudes, is defined by
 \begin{equation}\label{a1}
   R(\mathcal{B}_c\to \mathcal{B}K^0_{S,L})\equiv\frac{\Gamma(\mathcal{B}_c\rightarrow \mathcal{B}K_S^0) -\Gamma(\mathcal{B}_c\rightarrow \mathcal{ B}K_L^0)}{\Gamma(\mathcal{B}_c\rightarrow \mathcal{B}K_S^0) + \Gamma(\mathcal{B}_c\rightarrow \mathcal{B}K_L^0)}.
 \end{equation}
If the ratio between the DCS and CF amplitudes is defined as
\begin{equation}\label{a3}
 \mathcal{A}(\mathcal{B}_c\to \mathcal{B}K^0)/ \mathcal{A}(\mathcal{B}_c\to\mathcal{B}\overline K^0) \equiv
  r_f e^{i(\phi+\delta_f)},
\end{equation}
with the magnitude $r_f$, the relative strong phase $\delta_f$, and the weak phase
$\phi\equiv Arg\left[-V_{cd}^{*}V_{us}/V_{cs}^{*}V_{ud} \right]
=(-6.2\pm 0.4)\times 10^{-4}$,
the $K^0_S-K^0_L$ asymmetry is reduced to be \cite{Wang:2017ksn}
\begin{equation}\label{x4}
   R(\mathcal{B}_c\to \mathcal{B}K^0_{S,L}) \simeq -2r_f\cos\delta_f.
\end{equation}
The time-dependent $CP$ asymmetry in the decay chain of $\mathcal{B}_c\to \mathcal{B}K(t)(\to \pi^+\pi^-)$ is defined as
\begin{equation}\label{m1}
A_{CP}(t) \equiv\frac{\Gamma_{\pi\pi}(t)-\overline
\Gamma_{\pi\pi}(t)}{\Gamma_{\pi\pi}(t)+\overline\Gamma_{\pi\pi}(t)},
\end{equation}
with $  \Gamma_{\pi\pi}(t)\equiv\Gamma(\mathcal{B}_c\to \mathcal{B}K(t)(\to \pi^{+}\pi^{-}))$
and $\overline\Gamma_{\pi\pi}(t)\equiv\Gamma(\overline {\mathcal{B}}_c\to \overline {\mathcal{B}}K(t)
(\to \pi^{+}\pi^{-}))$.
The intermediate state $K(t)$ is recognized as a time-evolved neutral kaon $K^0(t)$ or $\overline K^0(t)$, and $t$ is the time interval between the charmed baryon decay and the neutral kaon decay in the kaon rest frame  \cite{Grossman:2011zk,Yu:2017oky}.
As pointed out in \cite{Yu:2017oky}, there exist three $CP$-violation effects, i.e., the indirect $CP$ violation in $K^{0}-\overline K^{0}$ mixing $A_{CP}^{\overline K^0}$, the direct $CP$ asymmetry in charm decays $A_{CP}^{\rm dir}$, and the effect from the interference between two tree (CF and DCS) amplitudes with neutral kaon mixing $A_{CP}^{\rm int}$,
\begin{equation}\label{eq:KSAcp}
A_{CP}(t)\simeq\big[A_{CP}^{\overline K^0}(t)+A_{CP}^{\text{dir}}(t)+A_{CP}^{\text{int}}(t)\big]/D(t),
\end{equation}
in which
\begin{align}
A_{CP}^{\overline K^0}(t)
&=  2\mathcal{R}e(\epsilon)e^{-\Gamma_St}-2e^{-\Gamma t}
\big(\mathcal{R}e(\epsilon)\cos(\Delta mt)+\mathcal{I}m(\epsilon)\sin(\Delta mt)\big),\\\label{a2}
A_{CP}^{\text{dir}}(t)&=2e^{-\Gamma_St}\,r_f\sin\delta_f\sin\phi,\\\label{a4}
A_{CP}^{\text{int}}(t)
&= -4\,r_f\cos\phi\sin\delta_f\big(\mathcal{I}m(\epsilon)e^{-\Gamma_St}-e^{-\Gamma t}
(\mathcal{I}m(\epsilon)\cos(\Delta mt)-\mathcal{R}e(\epsilon)\sin(\Delta mt))\big),\\
D(t)&= e^{-\Gamma_St}(1-2\,r_f\cos\delta_f\cos\phi)+e^{-\Gamma_Lt}|\epsilon|^2,
 \end{align}
with the parameter $\epsilon$ characterizing the indirect $CP$ asymmetry in the $K^0-\overline K^0$ mixing, the mass $m_{S}$ ($m_L$) and the width $\Gamma_{S}$ ($\Gamma_L$) of the $K^0_S$ ($K^0_L$) meson and $\Gamma\equiv(\Gamma_S+\Gamma_L)/2$, $\Delta m\equiv m_L-m_S$.

In the $\mathcal{B}_c\to \mathcal{B}_8P$ and $\mathcal{B}_c\to \mathcal{B}_{10}P$ decays, we can define seven $K^0_S-K^0_L$ asymmetries which are associated with the decay modes of $\Lambda^+_c\to pK^0_{S,L}$, $\Xi^+_c\to \Sigma^+K^0_{S,L}$, $\Xi^0_c\to \Sigma^0K^0_{S,L}$, $\Xi^0_c\to \Lambda^0K^0_{S,L}$, $\Lambda^+_c\to \Delta^+K^0_{S,L}$, $\Xi^+_c\to \Sigma^{*+}K^0_{S,L}$, $\Xi^0_c\to \Sigma^{*0}K^0_{S,L}$.
Let us analyze the relation between $R(\Lambda^+_c\to pK^0_{S,L})$ and $R(\Xi^+_c\to \Sigma^+K^0_{S,L})$ in the $U$-spin symmetry.
The ratio of decay amplitudes of $\Lambda^+_c\to pK^0$ and $\Lambda^+_c\to p\overline K^0$ is
\begin{align}\label{eqsum1}
   \frac{\mathcal{A}(\Lambda^+_c\to pK^0)}{\mathcal{A}(\Lambda^+_c\to p\overline K^0)} = -\frac{V^*_{cd}V_{us}}{V^*_{cs}V_{ud}}\frac{c+d+g}{a+c+e}
 =r_{p}e^{i(\phi+\delta_{p})}.
\end{align}
The ratio of decay amplitudes of $\Xi^+_c\to \Sigma^+K^0$ and $\Xi^+_c\to \Sigma^+\overline K^0$ is
\begin{align}\label{eqsum2}
   \frac{\mathcal{A}(\Xi^+_c\to \Sigma^+K^0)}{\mathcal{A}(\Xi^+_c\to \Sigma^+\overline K^0)} = -\frac{V^*_{cd}V_{us}}{V^*_{cs}V_{ud}}\frac{a+c+e}{c+d+g}
    =r_{\Sigma^+}e^{i(\phi+\delta_{\Sigma^+})}.
\end{align}
Eqs.~\eqref{eqsum1} and \eqref{eqsum2} show that the magnitude of ratios $r_{p}$ and $r_{\Sigma^+}$ and strong phases $\delta_p$ and $\delta_{\Sigma^+}$ have following relations in the $U$-spin limit (under the approximation of Eq.~\eqref{gam}):
\begin{align}\label{eqsum3}
r_{p}/\tan^2\theta=\tan^2\theta/r_{\Sigma^+}, \qquad \delta_{p}=-\delta_{\Sigma^+},
\end{align}
For convenience, we define $r_{p}/\tan^2\theta=\tan^2\theta/r_{\Sigma^+}=-\hat{r}$, $\delta_{p}=-\delta_{\Sigma^+}=\hat{\delta}$.
Then the $K^0_S-K^0_L$ asymmetries in $\Lambda^+_c\to pK^0_{S,L}$ and $\Xi^+_c\to \Sigma^+K^0_{S,L}$ modes can be written as
\begin{align}
R(\Lambda^+_c\to pK^0_{S,L})\simeq 2\tan^2\theta\,\hat{r}\cos\hat{\delta},\qquad R(\Xi^+_c\to \Sigma^+K^0_{S,L})\simeq 2\tan^2\theta\cos\hat{\delta}/\hat{r}.
\end{align}
The following relation between $R(\Lambda^+_c\to pK^0_{S,L})$ and $R(\Xi^+_c\to \Sigma^+K^0_{S,L})$ is gotten:
\begin{equation}\label{f1}
  R(\Lambda^+_c\to pK^0_{S,L})\times R(\Xi^+_c\to \Sigma^+K^0_{S,L}) = 4\tan^4\theta\cos^2\hat{\delta}\leq4\tan^4\theta \approx 1\times 10^{-2}.
\end{equation}

According to Eqs.~\eqref{a2} and \eqref{a4}, the direct $CP$ asymmetry $A_{CP}^{\rm dir}$ and the interference between charm decay and neutral kaon mixing $A_{CP}^{\rm int}$ are proportional to $\sin\delta_f$.
Since the relative strong phase between DCS and CF amplitudes $\delta_f$ is opposite in $\Lambda^+_c\to pK^0_{S}$ and $\Xi^+_c\to \Sigma^+K^0_{S}$ modes, $A_{CP}^{\rm dir}$ and $A_{CP}^{\rm int}$
have opposite sign too.
Furthermore, the magnitudes of $K^0_S-K^0_L$ asymmetries and $CP$ asymmetries in $\Lambda^+_c\to pK^0_{S,L}$ and $\Xi^+_c\to \Sigma^+K^0_{S,L}$ modes have following relations:
\begin{enumerate}
  \item If both $R(\Lambda^+_c\to pK^0_{S,L})$ and $R(\Xi^+_c\to \Sigma^+K^0_{S,L})$ are large, the strong phase $\hat{\delta}$ is close to zero. The $CP$ asymmetries $A_{CP}^{\rm dir}$ and $A_{CP}^{\rm int}$ are small in $\Lambda^+_c\to pK^0_{S}$ and $\Xi^+_c\to \Sigma^+K^0_{S}$ decays.
  \item If both $R(\Lambda^+_c\to pK^0_{S,L})$ and $R(\Xi^+_c\to \Sigma^+K^0_{S,L})$ are small, the strong phase $\hat{\delta}$ is close to $\pi/2$. The $CP$ asymmetries $A_{CP}^{\rm dir}$ and $A_{CP}^{\rm int}$ are large in $\Lambda^+_c\to pK^0_{S}$ and $\Xi^+_c\to \Sigma^+K^0_{S}$ decays.
  \item If $R(\Lambda^+_c\to pK^0_{S,L})$ is large while $R(\Xi^+_c\to \Sigma^+K^0_{S,L})$ is small, the parameter $\hat{r}$ is large. The $CP$ asymmetries $A_{CP}^{\rm dir}$ and $A_{CP}^{\rm int}$ in $\Lambda^+_c\to pK^0_{S}$ decay are large and the ones in $\Xi^+_c\to \Sigma^+K^0_{S}$ decay are small.
  \item If $R(\Lambda^+_c\to pK^0_{S,L})$ is small while $R(\Xi^+_c\to \Sigma^+K^0_{S,L})$ is large, the parameter $\hat{r}$ is small. The $CP$ asymmetries $A_{CP}^{\rm dir}$ and $A_{CP}^{\rm int}$ in $\Lambda^+_c\to pK^0_{S}$ decay are small and the ones in $\Xi^+_c\to \Sigma^+K^0_{S}$ decay are large.
\end{enumerate}

Let us take a closer look on Eqs.~\eqref{eqsum1} and \eqref{eqsum2}. The decay modes $\Lambda^+_c\to pK^0$ and $\Xi^+_c\to \Sigma^+\overline K^0$ are connected by a complete interchange of $d$ and $s$ quarks in initial and final states: $\Lambda^+_c \leftrightarrow \Xi^+_c$, $p \leftrightarrow \Sigma^+$, $K^0\leftrightarrow \overline K^0$. The decay amplitudes of $\Lambda^+_c\to pK^0$ and $\Xi^+_c\to \Sigma^+\overline K^0$ are associated with an interchange of the CKM matrix elements: $V^{*}_{cd}V_{us} \leftrightarrow V^{*}_{cs}V_{ud} $.
The same situation appears in $\Lambda^+_c\to p\overline K^0$ and $\Xi^+_c\to \Sigma^+ K^0$ modes.
In fact, it is an universal law that if two decay modes connected by the interchange of $d \leftrightarrow s$ in the initial and final states,
 their decay amplitudes are the same under the flavor $U$-spin symmetry except for an interchange of the CKM matrix elements. The detailed analysis can be found in our previous work \cite{Wang:2019dls}.
Even the decay amplitudes are not written down, Eq.~\eqref{eqsum3} can still be obtained.
Eq.~\eqref{eqsum3} is valid for any charmed decay modes involving $K^0_{S,L}$ if other initial- and final-state particles are connected by a complete interchange of $d$ and $s$ quarks, no matter the charmed meson decays, singly and doubly charmed baryon decays, or two- and multi-body decays. All the analysis on $\Lambda^+_c\to p K^0_{S,L}$ and $\Xi^+_c\to \Sigma^+ K^0_{S,L}$ can be applied to the modes that satisfy this condition. As examples, one can check decay modes such as $D^+\to K^0_{S,L}\pi^+$ and $D^+_s\to K^0_{S,L}K^+$, $\Lambda_{c}^{+}\to \Delta^+K^0_{S,L}$ and $\Xi_c^+\to \Sigma^{*+}K^0_{S,L}$, $\Xi_{cc}^+\to \Lambda^+_cK^0_{S,L}$ and $\Omega^+_{cc}\to \Xi^+_cK^0_{S,L}$, $\Xi_{cc}^+\to \Sigma^+_cK^0_{S,L}$ and $\Omega^+_{cc}\to \Xi^{*+}_cK^0_{S,L}$, and so on.

\subsection{$U$-spin breaking}\label{re1}
As is well known the $SU(3)$ breaking effects are significantly large in the charmed meson decays \cite{Li:2012cfa,Li:2013xsa}.
It deserves to investigate the $SU(3)$ breaking effects in charmed baryon decays. In charm decays, $U$-spin breaking is usually studied.
A perturbative method to deal with $U$-spin breaking was proposed in \cite{Gronau:2013xba,Gronau:2015rda}. In this method, the corrections of arbitrary order to decay amplitude $\langle f|\mathcal{H}_{\rm eff}|i\rangle$ are obtained by introducing an $s-d$ spurion mass operator $M_{U\rm brk}$ into the Hamiltonian and initial and final states. It is the $U_3=0$ component in $U$-spin triplet.
Using the $s-d$ spurion mass operator, the author of \cite{Gronau:2013xba,Gronau:2015rda} derives the $n$-th order $U$-spin breaking corrections for $D^0\to K^-\pi^+$, $D^0\to K^+K^-$,  $D^0\to \pi^+\pi^-$ and $D^0\to K^+\pi^-$ decays.
In this subsection, we try to apply this perturbative method to charmed baryon decays to analyze the singly Cabibbo-suppressed modes $\Lambda_c^+\to \Sigma^+K^{*0}$ and $\Xi_c^+\to p\overline{K}^{*0}$.

Under the $U$-spin symmetry limit, the decay amplitudes of $\Lambda_c^+\to \Sigma^+K^{*0}$ and $\Xi_c^+\to p\overline{K}^{*0}$ are equal,
\begin{equation}
\langle\Sigma^+K^{*0}|\mathcal{H}_{\rm eff}^{\rm SCS}|\Lambda_c^+\rangle^{(0)}=\langle p\overline{K}^{*0}|\mathcal{H}_{\rm eff}^{\rm SCS}|\Xi_c^+\rangle^{(0)} = \mathcal{A},
\end{equation}
as shown in Eq.~(\ref{Uspin9}).
The first order $U$-spin breaking is obtained by multiplying an $s-d$ spurion mass operator $M_{U\rm brk}\propto(\bar ss)-(\bar dd)$ with Hamiltonian and initial and final states,
\begin{equation}
 \langle f|\mathcal{H}_{\rm eff}^{\rm SCS}|i\rangle^{(1)}=\langle f|\mathcal{H}_{\rm eff}^{\rm SCS}M_{U\rm brk}|i\rangle+\langle f|\mathcal{H}_{\rm eff}^{\rm SCS}|i M_{U\rm brk}\rangle+\langle M_{U\rm brk}f|\mathcal{H}_{\rm eff}^{\rm SCS}|i\rangle.
\end{equation}
The effective Hamiltonian for SCS decays at first order has an additional penguin term $P_{s+d}$ due to the $s-d$ mass difference \cite{Gronau:2013xba,Gronau:2015rda}
\begin{equation}
 \mathcal{H}_{\rm eff}^{\rm SCS}M_{U\rm brk}=\mathcal{H}_{W}^{\rm SCS}M_{U\rm brk}+P_{s+d}.
\end{equation}
Hence the first order correction of $U$-spin breaking is rewritten as
\begin{equation}\label{eff}
 \langle f|\mathcal{H}_{\rm eff}^{\rm SCS}|i\rangle^{(1)}=\langle f|\mathcal{H}_{W}^{\rm SCS}M_{U\rm brk}|i\rangle+\langle f|P_{s+d}|i\rangle+\langle f|\mathcal{H}_{\rm eff}^{\rm SCS}|i M_{U\rm brk}\rangle+\langle M_{U\rm brk}f|\mathcal{H}_{\rm eff}^{\rm SCS}|i\rangle,
\end{equation}
in which
\begin{align}
 |\mathcal{H}_{\rm W}^{\rm SCS}\rangle=\sqrt2|1,0\rangle,\qquad
 |M_{U\rm brk}\rangle=|1,0\rangle, \qquad |P_{s+d}\rangle=|0,0\rangle.
\end{align}
For $\Lambda_c^+\to \Sigma^+K^{*0}$ and $\Xi_c^+\to p\overline{K}^{*0}$ decays,
\begin{align}
 |i\rangle=|\frac{1}{2},\pm\frac{1}{2}\rangle,\qquad
 |f\rangle=\frac{1}{\sqrt3}|\frac{3}{2},\pm\frac{1}{2}\rangle\pm\sqrt{\frac{2}{3}}|\frac{1}{2},\pm\frac{1}{2}\rangle.
\end{align}
According to the coupling law of angular momenta and the following property of Clebsch-Gordan coefficients,
\begin{equation}\label{property}
 \langle j_1m_1j_2m_2|j_3m_3\rangle=(-1)^{j_1+j_2-j_3}\langle j_1-m_1\,j_2-m_2|j_3-m_3\rangle,
\end{equation}
the relation between the decay amplitudes of $\Lambda_c^+\to \Sigma^+K^{*0}$ and $\Xi_c^+\to p\overline{K}^{*0}$ decays at first order correction of $U$-spin breaking can be derived. For instance, the expressions of the first term in Eq.~\eqref{eff} in $\Lambda_c^+\to \Sigma^+K^{*0}$ and $\Xi_c^+\to p\overline{K}^{*0}$ modes are written as
\begin{align}
 \langle\Sigma^+K^{*0}|\mathcal{H}_W^{\rm SCS}M_{U\rm brk}|\Lambda_c^+\rangle&=\langle\frac{1}{\sqrt3}(\frac{3}{2},\frac{1}{2})
 +\sqrt{\frac{2}{3}}(\frac{1}{2},\frac{1}{2})|\frac{2}{\sqrt3}(2,0)-\sqrt{\frac{2}{3}}(0,0)|\frac{1}{2},\frac{1}{2}\rangle\nonumber\\
 &=-\frac{2\sqrt2}{3\sqrt5}\mathcal{A}_{\frac{3}{2}}-\frac{2}{3}\mathcal{A}_{\frac{1}{2}},
\end{align}
\begin{align}
 \langle p\overline{K}^{*0}|\mathcal{H}_W^{\rm SCS}M_{U\rm brk}|\Xi_c^+\rangle&=\langle\frac{1}{\sqrt3}(\frac{3}{2},-\frac{1}{2})
 -\sqrt{\frac{2}{3}}(\frac{1}{2},-\frac{1}{2})|\frac{2}{\sqrt3}(2,0)-\sqrt{\frac{2}{3}}(0,0)|\frac{1}{2},-\frac{1}{2}\rangle\nonumber\\
 &=\frac{2\sqrt2}{3\sqrt5}\mathcal{A}_{\frac{3}{2}}+\frac{2}{3}\mathcal{A}_{\frac{1}{2}}.
\end{align}
One can find $\langle f|\mathcal{H}_W^{\rm SCS}M_{U\rm brk}|i\rangle$ term in $\Lambda_c^+\to \Sigma^+K^{*0}$ and $\Xi_c^+\to p\overline{K}^{*0}$  has opposite sign.
Similar conclusions are also deduced in other terms of Eq.~\eqref{eff}. Thereby,
the relation between ratios of the first order $U$-spin breaking amplitude and the $U$-spin symmetry amplitude of these two decay modes is
\begin{equation}
\frac{\langle\Sigma^+K^{*0}|\mathcal{H}_{\rm eff}^{\rm SCS}|\Lambda_c^+\rangle^{(1)}}{\langle\Sigma^+K^{*0}|\mathcal{H}_{\rm eff}^{\rm SCS}|\Lambda_c^+\rangle^{(0)}}
=-\frac{\langle p\overline{K}^{*0}|\mathcal{H}_{\rm eff}^{\rm SCS}|\Xi_c^+\rangle^{(1)}}{\langle p\overline{K}^{*0}|\mathcal{H}_{\rm eff}^{\rm SCS}|\Xi_c^+\rangle^{(0)}}\equiv\varepsilon_{\mathcal{B}}.
\end{equation}
So the amplitudes of $\Lambda_c^+\to \Sigma^+K^{*0}$ and $\Xi_c^+\to p\overline{K}^{*0}$ decays  in the first-order $U$-spin breaking are
\begin{align}
\mathcal{A}(\Lambda_c^+\to\Sigma^+K^{*0})=V_{\rm CKM}A\,(1+\varepsilon_{\mathcal{B}}),\qquad
\mathcal{A}(\Xi_c^+\to p\overline{K}^{*0})=V_{\rm CKM}A\,(1-\varepsilon_{\mathcal{B}}),
\end{align}
in which the $V_{\rm CKM}$ involves  $V_{cs}V_{us}\simeq\sin\theta\cos\theta$ and $V_{cd}V_{ud}\simeq-\sin\theta\cos\theta$.
Neglecting the high order contributions, the ratio of $\mathcal{A}(\Lambda_c^+\to\Sigma^+K^{*0})$ and $\mathcal{A}(\Xi_c^+\to p\overline{K}^{*0})$ is
\begin{equation}\label{b}
 \frac{\mathcal{A}(\Lambda_c^+\to\Sigma^+K^{*0})}{\mathcal{A}(\Xi_c^+\to p\overline{K}^{*0})}=-\frac{1+\varepsilon_{\mathcal{B}}}{1-\varepsilon_{\mathcal{B}}}
 =-(1+2\mathcal{R}e(\varepsilon_{\mathcal{B}}))+\mathcal{O}(\varepsilon_{\mathcal{B}}^2).
\end{equation}

With the experimental data of the branching fractions \cite{Li:2019atu}
\begin{align}
 \mathcal{B}r(\Xi_c^+\to\Xi^-\pi^+\pi^+)=(2.86\pm1.21\pm0.38)\%, \quad \mathcal{B}r(\Xi_c^+\to pK^-\pi^+)=(0.45\pm0.21\pm0.07)\%,
\end{align}
and the ratios \cite{Tanabashi:2018oca,Link:2001rn},
\begin{align}
\frac{\mathcal{B}r(\Xi_c^+\to p\overline{K}^{*0})}{\mathcal{B}r(\Xi_c^+\to\Xi^-\pi^+\pi^+)}=0.116\pm0.030,\qquad\frac{\mathcal{B}r(\Xi_c^+\to p\overline{K}^{*0})}{\mathcal{B}r(\Xi_c^+\to pK^-\pi^+)}=0.54\pm0.10,
\end{align}
we get two different branching fractions of $\Xi_c^+\to p\overline{K}^{*0}$:
\begin{align}
 \mathcal{B}r(\Xi_c^+\to p\overline{K}^{*0})^{(1)}=(3.32\pm1.70)\times10^{-3}, \qquad \mathcal{B}r(\Xi_c^+\to p\overline{K}^{*0})^{(2)}=(2.43\pm1.27)\times10^{-3}.
\end{align}
The average of them can be calculated through the following formulas
\begin{align}
 \bar\zeta=\sum\limits_{i=1}^nw_iY_i/\sum\limits_{i=1}^nw_i, \qquad \sigma\bar\zeta=\sqrt{1/\sum\limits_{i=1}^nw_i},
\end{align}
where $Y_i$ is the central value of each branching fraction with error $\sigma_i$ and the weight function $w_i$ is $1/\sigma_i^2$.
Then the branching fraction of $\Xi_c^+\to p\overline{K}^{*0}$ is
\begin{equation}
 \mathcal{B}r(\Xi_c^+\to p\overline{K}^{*0}) = (2.75\pm 1.02)\times 10^{-3}.
\end{equation}
The $U$-spin breaking parameter $\mathcal{R}e(\varepsilon_{\mathcal{B}})$ in $\Lambda_c^+\to \Sigma^+K^{*0}$ and $\Xi_c^+\to p\overline{K}^{*0}$ modes is extracted to be
\begin{equation}
 \mathcal{R}e(\varepsilon_{\mathcal{B}})=0.53\pm0.24.
\end{equation}
Compared to the ratio between the branching fractions of $\Lambda_c^+\to\Sigma^+\pi^0$ and $\Lambda_c^+\to\Sigma^0\pi^+$ decays in Eq.~\eqref{ib}, one can find the $U$-spin breaking is much larger than the $I$-spin breaking.
If the $U$-spin breaking in $\Lambda_c^+\to \Sigma^+K^{*0}$ and $\Xi_c^+\to p\overline{K}^{*0}$ decays is normal, i.e., no more than $30\%$, the parameter $\mathcal{R}e(\varepsilon_{\mathcal{B}})$ is smaller than $15\%$ since there is a factor $2$ in Eq.~\eqref{b}. $\mathcal{R}e(\varepsilon_{\mathcal{B}})$ extracted from data, at least its central value, is much larger than $15\%$. We can regard the abnormal $U$-spin breaking as an anomaly.
To confirm the anomaly, more accurate data are required. The large $U$-spin breaking in $\Lambda_c^+\to \Sigma^+K^{*0}$ and $\Xi_c^+\to p\overline{K}^{*0}$ decays is very interesting because similar anomaly is also found in the SCS charmed meson decay modes $D^0\to K^+K^-$ and $D^0\to \pi^+\pi^-$ \cite{Tanabashi:2018oca},
\begin{align}
\mathcal{B}r(D^0\to K^+K^-)= (4.08\pm 0.06)\times 10^{-3},\qquad \mathcal{B}r(D^0\to \pi^+\pi^-) = (1.445\pm 0.024)\times 10^{-3},
\end{align}
and
\begin{align}
\mathcal{B}r(D^0\to K^+K^-)/\mathcal{B}r(D^0\to \pi^+\pi^-) = 2.80\pm 0.02.
\end{align}
The $U$-spin breaking parameter $\mathcal{R}e(\varepsilon_D)$ is given by \cite{Gronau:2015rda},
\begin{align}
 \mathcal{R}e(\varepsilon_D)=0.310\pm0.006.
\end{align}
It is plausible that $U$-spin breaking in the singly Cabibbo-suppressed transitions is larger than the Cabibbo-favored and doubly Cabibbo-suppressed transitions and some non-pertubative dynamics enhance the $U$-spin breaking in both charmed meson and baryon decays.

\section{Summary}\label{sum}
In summary, we study the two-body non-leptonic decays of charmed baryons based on the flavor $SU(3)$ symmetry.
Hundreds of $I$-, $U$- and $V$-spin relations between different decay channels of charmed baryons are found. Some of them can be used to test the breaking of $I$-, $U$- and $V$-spins.
Using these amplitude relations, some branching fractions of charmed baryon decays are predicted, which could provide guides for the future experiments.
Several $U$-spin relations for $K^0_S-K^0_L$ asymmetries and $CP$ asymmetries in the $\mathcal{B}_c\to \mathcal{B}K^0_{S,L}$ modes are proposed.
And a possible abnormal $U$-spin breaking is found in $\Lambda_c^+\to \Sigma^+K^{*0}$ and $\Xi_c^+\to p\overline{K}^{*0}$ modes.

\begin{acknowledgments}
We are grateful to Cheng-Ping Shen, Pei-Rong Li and Hua-Yu Jiang for useful discussions.
This work was supported in part by  the National Natural Science Foundation of China under
Grants No. U1732101, 11975112, by Gansu Natural Science Fund under grant No.18JR3RA265, and by the Fundamental Research Funds for the
Central Universities under Grant No. lzujbky-2018-it33 and lzujbky-2019-55.
\end{acknowledgments}

\appendix

\section{Decay amplitudes}\label{amp}

In this Appendix, we list the decay amplitudes of all $\mathcal{B}_c\rightarrow \mathcal{B}_8P$, $\mathcal{B}_c\rightarrow \mathcal{B}_{10}P$, $\mathcal{B}_c\rightarrow \mathcal{B}_8V$ and $\mathcal{B}_c\rightarrow \mathcal{B}_{10}V$ modes, see Tables.~\ref{tab:amp1} $\sim$ \ref{tab:amp4}.

\begin{table*}[htp]
\caption{Decay amplitudes of $\mathcal{B}_c\rightarrow \mathcal{B}_8P$ modes in the $SU(3)_F$ limit, in which mod $s_1$ and $s^2_1$ are used to label the singly Cabibbo-suppressed and doubly Cabibbo-suppressed modes since the SCS amplitudes are proportional to $\sin \theta$ (mod $s_1$) and the DCS amplitudes are proportional to $\sin^2 \theta$ (mod $s^2_1$).}\label{tab:amp1}
\begin{ruledtabular}
\footnotesize\begin{tabular}{cc|cc}
\textbf{Mode} & \textbf{Decay amplitude} & $\Xi_c^+\to\Lambda^0K^+$ & $\frac{1}{\sqrt6}(a-2b+c+e-2f-2g)$ \\ \hline
$\Lambda_c^+\to\Lambda^0\pi^+$  & $\frac{1}{\sqrt6}(a+b-2c+e+f+g)$ & $\Xi_c^+\to p\pi^0$ & $\frac{1}{\sqrt2}(b-d+f)$ \\ \hline
$\Lambda_c^+\to\Sigma^0\pi^+$ & $\frac{1}{\sqrt2}(a-b+e-f-g)$ & $\Xi_c^+\to n\pi^+$ & $(b+d+f)$ \\ \hline
$\Lambda_c^+\to\Sigma^+\pi^0$ & $\frac{1}{\sqrt2}(-a+b-e+f+g)$ &  \textbf{Mode} & \textbf{Decay amplitude} (mod $s_1$) \\ \hline
$\Lambda_c^+\to\Xi^0 K^+$ & $(b+d+f)$ & $\Lambda_c^+\to\Lambda^0 K^+$ & $\frac{1}{\sqrt6}(a-2b-2c-3d+e-2f+g)$\\ \hline
$\Lambda_c^+\to p\overline{K}^0$ & $(a+c+e)$ &  $\Lambda_c^+\to\Sigma^0 K^+$ & $\frac{1}{\sqrt2}(a+d+e-g)$\\ \hline
$\Lambda_c^+\to\Sigma^+\eta$ & \makecell{$\frac{1}{\sqrt6}\cos\xi(a+b-2d+e+f-g)$\\$-\frac{1}{\sqrt3}\sin\xi(a+b+d+e+f-g+3h+3r)$} & $\Lambda_c^+\to p\eta$ & \makecell{$\frac{1}{\sqrt6}\cos\xi(-2a+b-3c-2d-2e+f-g)$\\$-\frac{1}{\sqrt3}\sin\xi(a+b+d+e+f-g+3h+3r)$} \\ \hline
$\Lambda_c^+\to\Sigma^+\eta^\prime$ & \makecell{$\frac{1}{\sqrt6}\sin\xi(a+b-2d+e+f-g)$\\$+\frac{1}{\sqrt3}\cos\xi(a+b+d+e+f-g+3h+3r)$} & $\Lambda_c^+\to p\eta^\prime$ & \makecell{$\frac{1}{\sqrt6}\sin\xi(-2a+b-3c-2d-2e+f-g)$\\$+\frac{1}{\sqrt3}\cos\xi(a+b+d+e+f-g+3h+3r)$}\\ \hline
$\Xi_c^0\to\Xi^0\pi^0$ & $\frac{1}{\sqrt2}(-a+d+e-g)$ & $\Lambda_c^+\to p \pi^0$ & $\frac{1}{\sqrt2}(b+c+f+g)$\\ \hline
$\Xi_c^0\to\Lambda^0\overline{K}^0$ & $\frac{1}{\sqrt6}(-2a+b+c+2e-f-g)$ & $\Lambda_c^+\to n \pi^+$ & $(b-c+f+g)$\\ \hline
$\Xi_c^0\to\Xi^-\pi^+$ & $(a+c-e)$ & $\Lambda_c^+\to\Sigma^+ K^0$ & $(a-d+e-g)$ \\ \hline
$\Xi_c^0\to\Xi^0\eta$ & \makecell{$\frac{1}{\sqrt6}\cos\xi(a-2b+d-e+2f+g)$\\$-\frac{1}{\sqrt3}\sin\xi(a+b+d-e-f+g+3h-3r)$} & $\Xi_c^0\to\Sigma^0\eta$ & \makecell{$\frac{1}{2\sqrt{3}}\cos\xi(a+b-3c+d-e-f-2g)$\\$-\frac{1}{\sqrt6}\sin\xi(a+b+d-e-f+g+3h-3r)$}\\ \hline
$\Xi_c^0\to\Xi^0\eta^\prime$ & \makecell{$\frac{1}{\sqrt6}\sin\xi(a-2b+d-e+2f+g)$\\$+\frac{1}{\sqrt3}\cos\xi(a+b+d-e-f+g+3h-3r)$} & $\Xi_c^0\to\Sigma^0\eta^\prime$ & \makecell{$\frac{1}{2\sqrt{3}}\sin\xi(a+b-3c+d-e-f-2g)$\\$+\frac{1}{\sqrt6}\cos\xi(a+b+d-e-f+g+3h-3r)$}\\ \hline
$\Xi_c^0\to\Sigma^+ K^-$ & $(b+d-f)$ & $\Xi_c^0\to\Sigma^-\pi^+$ & $(-a-c+e)$\\ \hline
$\Xi_c^0\to\Sigma^0\overline{K}^0$ & $\frac{1}{\sqrt2}(-b+c+f+g)$ & $\Xi_c^0\to \Lambda^0\pi^0$ & $\frac{1}{2\sqrt{3}}(a+b+c-3d-e-f+2g)$\\ \hline
$\Xi_c^+\to\Xi^0\pi^+$ & $(-c-d+g)$ & $\Xi_c^0\to\Sigma^0\pi^0$ & $\frac{1}{2}(-a-b+c+d+e+f)$\\ \hline
$\Xi_c^+\to\Sigma^+\overline{K}^0$ & $(-c-d-g)$ & $\Xi_c^0\to n\overline{K}^0$ & $(-a+b+e-f-g)$\\ \hline
\textbf{Mode} & \textbf{Decay amplitude} (mod~$s_1^2$) & $\Xi_c^0\to\Sigma^+\pi^-$ & $(-b-d+f)$\\ \hline
$\Lambda_c^+\to p K^0$ & $(-c-d-g)$ &  $\Xi_c^0\to pK^-$ & $(b+d-f)$\\ \hline
$\Lambda_c^+\to n K^+$ & $(-c-d+g)$ & $\Xi_c^0\to\Xi^-K^+$ & $(a+c-e)$\\ \hline
$\Xi_c^0\to\Sigma^-K^+$ & $(-a-c+e)$ & $\Xi_c^0\to\Xi^0K^0$ & $(a-b-e+f+g)$\\ \hline
$\Xi_c^0\to \Lambda^0K^0$ & $\frac{1}{\sqrt6}(-a+2b-c+e-2f-2g)$ & $\Xi_c^+\to p\overline{K}^0$ & $(a-d+e-g)$\\ \hline
$\Xi_c^0\to\Sigma^0K^0$ & $\frac{1}{\sqrt2}(a-c-e)$ & $\Xi_c^+\to\Xi^0 K^+$ & $(b-c+f+g)$\\ \hline
$\Xi_c^0\to n\pi^0$ & $\frac{1}{\sqrt2}(b-d-f)$ & $\Xi_c^+\to\Lambda^0\pi^+$ & $\frac{1}{\sqrt6}(a+b+c+3d+e+f-2g)$\\ \hline
$\Xi_c^0\to p\pi^-$ & $(-b-d+f)$  & $\Xi_c^+\to\Sigma^0\pi^+$ & $\frac{1}{\sqrt2}(a-b-c-d+e-f)$ \\ \hline
$\Xi_c^0\to n\eta$ & \makecell{$\frac{1}{\sqrt6}\cos\xi(2a-b-d-2e+f+2g)$\\$+\frac{1}{\sqrt3}\sin\xi(a+b+d-e-f+g+3h-3r)$} & $\Xi_c^0\to\Lambda^0\eta$ & \makecell{$\frac{1}{2}\cos\xi(a+b-c-d-e-f)$\\$+\frac{1}{\sqrt{2}}\sin\xi(a+b+d-e-f+g+3h-3r)$}\\ \hline
$\Xi_c^0\to n\eta^\prime$ & \makecell{$\frac{1}{\sqrt6}\sin\xi(2a-b-d-2e+f+2g)$\\$-\frac{1}{\sqrt3}\cos\xi(a+b+d-e-f+g+3h-3r)$} & $\Xi_c^0\to\Lambda^0\eta^\prime$ & \makecell{$\frac{1}{2}\sin\xi(a+b-c-d-e-f)$\\$-\frac{1}{\sqrt{2}}\cos\xi(a+b+d-e-f+g+3h-3r)$}\\ \hline
$\Xi_c^+\to p\eta$ & \makecell{$\frac{1}{\sqrt6}\cos\xi(-2a+b+d-2e+f+2g)$\\$-\frac{1}{\sqrt3}\sin\xi(a+b+d+e+f-g+3h+3r)$} & $\Xi_c^+\to\Sigma^+\eta$ & \makecell{$\frac{1}{\sqrt6}\cos\xi(a+b+3c+d+e+f+2g)$\\$-\frac{1}{\sqrt3}\sin\xi(a+b+d+e+f-g+3h+3r)$} \\ \hline
$\Xi_c^+\to p\eta^\prime$ & \makecell{$\frac{1}{\sqrt6}\sin\xi(-2a+b+d-2e+f+2g)$\\$+\frac{1}{\sqrt3}\cos\xi(a+b+d+e+f-g+3h+3r)$} & $\Xi_c^+\to\Sigma^+\eta^\prime$ & \makecell{$\frac{1}{\sqrt6}\sin\xi(a+b+3c+d+e+f+2g)$\\$+\frac{1}{\sqrt3}\cos\xi(a+b+d+e+f-g+3h+3r)$}\\ \hline
$\Xi_c^+\to\Sigma^0 K^+$ & $\frac{1}{\sqrt2}(a-c+e)$ & $\Xi_c^+\to\Sigma^+\pi^0$ & $\frac{1}{\sqrt2}(-a+b-c-d-e+f)$\\ \hline
$\Xi_c^+\to\Sigma^+K^0$ & $(a+c+e)$ &&
\end{tabular}
\end{ruledtabular}
\end{table*}

\begin{table*}[htp]
\caption{Decay amplitudes of $\mathcal{B}_c\rightarrow \mathcal{B}_{10}P$ modes in the $SU(3)_F$ limit.}\label{tab:amp2}
\begin{ruledtabular}
\footnotesize\begin{tabular}{cc|cc}
\textbf{Mode} & \textbf{Decay amplitude}  &  $\Xi_c^0\to\Delta^-\pi^+$ & $(-\beta+\delta)$  \\
\hline
$\Lambda_c^+\to\Sigma^{*+}\pi^0$ & $-\frac{1}{\sqrt6}(2\alpha-\beta+2\gamma+\delta)$ & $\Xi_c^0\to\Sigma^{*-}K^+$ & $\frac{1}{\sqrt3}(-\beta+\delta)$ \\ \hline
$\Lambda_c^+\to\Sigma^{*0}\pi^+$ & $-\frac{1}{\sqrt6}(2\alpha-\beta+2\gamma+\delta)$& $\Xi_c^0\to\Delta^0\pi^0$ & $\frac{2}{\sqrt6}(-\gamma+\delta)$\\ \hline
$\Lambda_c^+\to\Sigma^{*+}\eta$ & \makecell{$\frac{1}{3\sqrt{2}}\cos\xi(2\alpha-\beta-2\gamma-3\delta)$\\$-\frac{1}{3}\sin\xi(\alpha+\beta-\gamma+3\lambda)$} & $\Xi_c^0\to\Delta^0\eta$ & \makecell{$\frac{\sqrt2}{3}\cos\xi(2\alpha-\beta+\gamma)$\\$+\frac{2}{3}\sin\xi(\alpha+\beta-\gamma+3\lambda)$} \\ \hline
$\Lambda_c^+\to\Sigma^{*+}\eta^\prime$ & \makecell{$\frac{1}{3\sqrt2}\sin\xi(2\alpha-\beta-2\gamma-3\delta)$\\$+\frac{2}{3}\cos\xi(\alpha+\beta-\gamma+3\lambda)$} & $\Xi_c^0\to\Delta^0\eta^\prime$ & \makecell{$\frac{\sqrt2}{3}\sin\xi(2\alpha-\beta+\gamma)$\\$-\frac{2}{3}\cos\xi(\alpha+\beta-\gamma+3\lambda)$}\\ \hline
$\Lambda_c^+\to\Delta^{++}K^-$ & $(\beta+\delta)$  & \textbf{Mode} & \textbf{Decay amplitude} (mod $s_1$)\\ \hline
$\Lambda_c^+\to\Xi^{*0} K^+$ & $\frac{1}{\sqrt3}(\beta-2\gamma-\delta)$ &  $\Lambda_c^+\to\Delta^+\pi^0$ & $\frac{2}{\sqrt6}(\alpha+\gamma+\delta)$\\ \hline
$\Lambda_c^+\to \Delta^+\overline{K}^0$ & $\frac{1}{\sqrt3}(\beta+\delta)$ & $\Lambda_c^+\to\Delta^0\pi^+$ & $\frac{1}{\sqrt3}(2\alpha-\beta+2\gamma+\delta)$ \\ \hline
$\Xi_c^0\to\Xi^{*0}\eta$ & \makecell{$\frac{1}{3\sqrt{2}}\cos\xi(-2\alpha+\beta-4\gamma+3\delta)$\\$+\frac{2}{3}\sin\xi(\alpha+\beta-\gamma+3\lambda)$} & $\Lambda_c^+\to\Delta^+\eta$ & \makecell{$-\frac{\sqrt2}{3}\cos\xi(\alpha+\beta-\gamma)$\\$-\frac{2}{3}\sin\xi(\alpha+\beta-\gamma+3\lambda)$}\\ \hline
$\Xi_c^0\to\Xi^{*0}\eta^\prime$ & \makecell{$\frac{1}{3\sqrt2}\sin\xi(-2\alpha+\beta-4\gamma+3\delta)$\\$-\frac{2 }{3}\cos\xi (\alpha+\beta-\gamma+3\lambda)$}& $\Lambda_c^+\to\Delta^+\eta^\prime$ & \makecell{$-\frac{\sqrt2}{3}\sin\xi(\alpha+\beta-\gamma)$\\$-\frac{2}{3}\cos\xi(\alpha+\beta-\gamma+3\lambda)$} \\ \hline
$\Xi_c^0\to\Sigma^{*0}\overline{K}^0$ & $\frac{1}{\sqrt6}(2\alpha-\beta+2\gamma-\delta)$ & $\Lambda_c^+\to\Sigma^{*+}K^0$ & $\frac{1}{\sqrt3}(2\alpha-\beta-\delta)$\\ \hline
$\Xi_c^0\to\Omega^-K^+$ & $(-\beta+\delta)$ & $\Lambda_c^+\to \Sigma^{*0}K^+$ & $\frac{1}{\sqrt6}(-2\alpha-\beta+2\gamma+\delta)$ \\ \hline
$\Xi_c^0\to\Xi^{*0}\pi^0$ & $\frac{1}{\sqrt6}(2\alpha-\beta+\delta)$ & $\Lambda_c^+\to\Delta^{++} \pi^-$ & $(-\beta-\delta)$\\ \hline
$\Xi_c^0\to\Sigma^{*+} K^-$ & $\frac{1}{\sqrt3}(-\beta+2\gamma-\delta)$ &  $\Xi_c^+\to\Sigma^{*+}\eta$ & \makecell{$\frac{1}{3\sqrt{2}}\cos\xi(-4\alpha-\beta-2\gamma-3\delta)$\\$-\frac{2}{3}\sin\xi(\alpha+\beta-\gamma+3\lambda)$} \\ \hline
$\Xi_c^0\to\Xi^{*-}\pi^+$ & $\frac{1}{\sqrt3}(-\beta+\delta)$ & $\Xi_c^+\to\Sigma^{*+}\eta^\prime$ & \makecell{$\frac{1}{3\sqrt{2}}\sin\xi(-4\alpha-\beta-2\gamma-3\delta)$\\$+\frac{2}{3}\cos\xi(\alpha+\beta-\gamma+3\lambda)$} \\ \hline
$\Xi_c^+\to\Sigma^{*+}\overline{K}^0$ & $\frac{2}{\sqrt3}\alpha$ &   $\Xi_c^+\to\Sigma^{*0}\pi^+$ & $\frac{1}{\sqrt6}(2\alpha+\beta-2\gamma-\delta)$\\ \hline
$\Xi_c^+\to\Xi^{*0}\pi^+$ & $-\frac{2}{\sqrt3}\alpha$ & $\Xi_c^+\to\Delta^+\overline{K}^0$ & $\frac{1}{\sqrt3}(-2\alpha+\beta+\delta)$ \\ \hline
\textbf{Mode} & \textbf{Decay amplitude} (mod $s_1^2$) &  $\Xi_c^+\to\Xi^{*0} K^+$ & $\frac{1}{\sqrt3}(-2\alpha+\beta-2\gamma-\delta)$ \\ \hline
$\Lambda_c^+\to\Delta^+K^0$ & $-\frac{2}{\sqrt3}\alpha$ &  $\Xi_c^+\to\Delta^{++}K^-$ & $(\beta+\delta)$ \\ \hline
$\Lambda_c^+\to\Delta^0 K^+$ & $\frac{2}{\sqrt3}\alpha$  &   $\Xi_c^+\to\Sigma^{*+}\pi^0$ & $\frac{1}{\sqrt6}(\beta-2\gamma-\delta)$ \\ \hline
$\Xi_c^+\to\Delta^{++}\pi^-$ & $(-\beta-\delta)$  &  $\Xi_c^0\to\Delta^0\overline{K}^0$ & $\frac{1}{\sqrt3}(-2\alpha+\beta-2\gamma+\delta)$\\ \hline
$\Xi_c^+\to\Sigma^{*+}K^0$ & $\frac{1}{\sqrt3}(-\beta-\delta)$ &  $\Xi_c^0\to\Xi^{*0} K^0$ & $\frac{1}{\sqrt3}(-2\alpha+\beta-2\gamma+\delta)$ \\ \hline
$\Xi_c^+\to\Delta^+\eta$ & \makecell{$\frac{\sqrt2}{3}\cos\xi(2\alpha-\beta+\gamma)$\\$+\frac{2}{3}\sin\xi(\alpha+\beta-\gamma+3\lambda)$} & $\Xi_c^0\to\Sigma^{*0}\eta$ & \makecell{$\frac{1}{6}\cos\xi(-2\alpha+\beta+2\gamma-3\delta)$\\$-\frac{2\sqrt2}{3}\sin\xi(\alpha+\beta-\gamma+3\lambda)$}\\ \hline
$\Xi_c^+\to\Delta^+\eta^\prime$ & \makecell{$\frac{\sqrt2}{3}\sin\xi(2\alpha-\beta+\gamma)$\\$-\frac{2}{3}\cos\xi(\alpha+\beta-\gamma+3\lambda)$} & $\Xi_c^0\to\Sigma^{*0}\eta^\prime$ & \makecell{$\frac{1}{6}\sin\xi(-2\alpha+\beta+2\gamma-3\delta)$\\$+\frac{2\sqrt2}{3}\cos\xi(\alpha+\beta-\gamma+3\lambda)$}\\ \hline
$\Xi_c^+\to\Delta^0\pi^+$ & $\frac{1}{\sqrt3}(-\beta+2\gamma+\delta)$ & $\Xi_c^0\to\Sigma^{*+}\pi^-$ & $\frac{1}{\sqrt3}(\beta-2\gamma+\delta)$\\ \hline
$\Xi_c^+\to\Delta^+\pi^0$ & $\frac{2}{\sqrt6}(\gamma+\delta)$ & $\Xi_c^0\to\Sigma^{*-}\pi^+$ & $\frac{2}{\sqrt3}(\beta-\delta)$\\ \hline
$\Xi_c^+\to\Sigma^{*0} K^+$ & $\frac{1}{\sqrt6}(2\alpha-\beta+2\gamma+\delta)$ & $\Xi_c^0\to\Xi^{*-}K^+$ & $\frac{2}{\sqrt3}(\beta-\delta)$\\ \hline
$\Xi_c^0\to\Sigma^{*0}K^0$ & $\frac{1}{\sqrt6}(2\alpha-\beta+2\gamma-\delta)$ & $\Xi_c^0\to\Delta^+K^-$ & $\frac{1}{\sqrt3}(\beta-2\gamma+\delta)$\\ \hline
$\Xi_c^0\to\Delta^+\pi^-$ & $\frac{1}{\sqrt3}(-\beta+2\gamma-\delta)$ & $\Xi_c^0\to\Sigma^{*0}\pi^0$ & $\frac{1}{2\sqrt{3}}(-2\alpha+\beta+2\gamma-3\delta)$
\end{tabular}
\end{ruledtabular}
\end{table*}

\begin{table*}[htp]
\caption{Decay amplitudes of $\mathcal{B}_c\rightarrow \mathcal{B}_8V$ modes in the $SU(3)_F$ limit.}\label{tab:amp3}
\begin{ruledtabular}
\footnotesize\begin{tabular}{cc|cc}
\textbf{Mode} & \textbf{Decay amplitude} &  $\Xi_c^+\to p\phi$ & $(-a^\prime-e^\prime+g^\prime-h^\prime-r^\prime)$\\
\hline
$\Lambda_c^+\to\Lambda^0\rho^+$ & $\frac{1}{\sqrt6}(a^\prime+b^\prime- 2c^\prime+e^\prime+f^\prime+g^\prime)$ &  $\Xi_c^+\to p\rho^0$ & $\frac{1}{\sqrt2}(b^\prime-d^\prime+f^\prime)$\\ \hline
$\Lambda_c^+\to\Sigma^0\rho^+$ & $\frac{1}{\sqrt2}(a^\prime-b^\prime+e^\prime-f^\prime-g^\prime)$ & $\Xi_c^+\to n\rho^+$ & $(b^\prime+d^\prime+f^\prime)$ \\ \hline
$\Lambda_c^+\to\Sigma^+\rho^0$ & $\frac{1}{\sqrt2}(-a^\prime+b^\prime-e^\prime+f^\prime+g^\prime)$ &  \textbf{Mode} & \textbf{Decay amplitude} (mod $s_1$)\\ \hline
$\Lambda_c^+\to\Sigma^+\phi$ & $(-d^\prime-h_1^\prime-h_2^\prime)$  & $\Lambda_c^+\to\Lambda^0 K^{*+}$ & $\frac{1}{\sqrt6}(a^\prime-2b^\prime-2c^\prime-3d^\prime+e^\prime-2f^\prime+g^\prime)$\\ \hline
$\Lambda_c^+\to p\overline{K}^{*0}$ & $(a^\prime+c^\prime+e^\prime)$ & $\Lambda_c^+\to\Sigma^0 K^{*+}$ & $\frac{1}{\sqrt2}(a^\prime+d^\prime+e^\prime-g^\prime)$ \\ \hline
$\Lambda_c^+\to\Xi^0 K^{*+}$ & $(b^\prime+d^\prime+f^\prime)$ & $\Lambda_c^+\to\Sigma^+ K^{*0}$ & $(a^\prime-d^\prime+e^\prime-g^\prime)$ \\ \hline
$\Lambda_c^+\to\Sigma^+\omega$ & $\frac{\sqrt2}{2}(a^\prime+b^\prime+e^\prime+f^\prime-g^\prime+2h^\prime+2r^\prime)$ & $\Lambda_c^+\to p\phi$ & $(-a^\prime-c^\prime-d^\prime-e^\prime-h^\prime-r^\prime)$\\ \hline
$\Xi_c^0\to\Xi^0\phi$ & $(-b^\prime+f^\prime-h^\prime+r^\prime)$ &  $\Lambda_c^+\to p \rho^0$ & $\frac{1}{\sqrt2}(b^\prime+c^\prime+f^\prime+g^\prime)$\\ \hline
$\Xi_c^0\to\Xi^-\rho^+$ & $(a^\prime+c^\prime-e^\prime)$ & $\Lambda_c^+\to n \rho^+$ & $(b^\prime-c^\prime+f^\prime+g^\prime)$ \\ \hline
$\Xi_c^0\to\Xi^0\rho^0$ & $\frac{1}{\sqrt2}(-a^\prime+d^\prime+e^\prime-g^\prime)$ & $\Lambda_c^+\to p\omega$ & $\frac{\sqrt2}{2}(b^\prime-c^\prime+f^\prime-g^\prime+2h^\prime+2r^\prime)$  \\ \hline
$\Xi_c^0\to\Xi^0\omega$ & $\frac{\sqrt2}{2}(a^\prime+d^\prime-e^\prime+g^\prime+2h^\prime-2r^\prime)$ & $\Xi_c^0\to\Sigma^0\omega$ & $\frac{1}{2}(a^\prime+b^\prime-c^\prime+d^\prime-e^\prime-f^\prime+2h^\prime-2r^\prime)$\\ \hline
$\Xi_c^0\to\Lambda^0\overline{K}^{*0}$ & $\frac{1}{\sqrt6}(-2a^\prime+b^\prime+c^\prime+2e^\prime-f^\prime-g^\prime)$ & $\Xi_c^0\to\Lambda^0\omega$ & $\frac{-1}{2\sqrt3}(a^\prime+b^\prime+c^\prime+3d^\prime-e^\prime-f^\prime+2g^\prime+6h^\prime-6r^\prime)$\\ \hline
$\Xi_c^0\to\Sigma^+ K^{*-}$ & $(b^\prime+d^\prime-f^\prime)$ &  $\Xi_c^0\to\Sigma^-\rho^+$ & $(-a^\prime-c^\prime+e^\prime)$\\ \hline
$\Xi_c^0\to\Sigma^0\overline{K}^{*0}$ & $\frac{1}{\sqrt2}(-b^\prime+c^\prime+f^\prime+g^\prime)$ & $\Xi_c^0\to \Lambda^0\rho^0$ & $\frac{1}{2\sqrt3}(a^\prime+b^\prime+c^\prime-3d^\prime-e^\prime-f^\prime+2g^\prime)$\\ \hline
$\Xi_c^+\to\Xi^0\rho^+$ & $(-c^\prime-d^\prime+g^\prime)$ &  $\Xi_c^0\to\Sigma^0\rho^0$ & $\frac{1}{2}(-a^\prime-b^\prime+c^\prime+d^\prime+e^\prime+f^\prime)$\\ \hline
$\Xi_c^+\to\Sigma^+\overline{K}^{*0}$ & $(-c^\prime-d^\prime-g^\prime)$ & $\Xi_c^0\to\Lambda^0\phi$ & $\frac{1}{\sqrt6}(2a^\prime+2b^\prime-c^\prime-2e^\prime-2f^\prime+g^\prime+3h^\prime-3r^\prime)$\\ \hline
\textbf{Mode} & \textbf{Decay amplitude} (mod $s_1^2$) & $\Xi_c^0\to\Sigma^0\phi$ & $\frac{\sqrt2}{2}(-c^\prime-g^\prime-h^\prime+r^\prime)$\\ \hline
$\Lambda_c^+\to p K^{*0}$ & $(-c^\prime-d^\prime-g^\prime)$ & $\Xi_c^0\to\Xi^-K^{*+}$ & $(a^\prime+c^\prime-e^\prime)$\\ \hline
$\Lambda_c^+\to n K^{*+}$ & $(-c^\prime-d^\prime+g^\prime)$ & $\Xi_c^0\to\Xi^0K^{*0}$ & $(a^\prime-b^\prime-e^\prime+f^\prime+g^\prime)$\\ \hline
$\Xi_c^0\to\Sigma^-K^{*+}$ & $(-a^\prime-c^\prime+e^\prime)$ & $\Xi_c^0\to\Sigma^+\rho^-$ & $(-b^\prime-d^\prime+f^\prime)$\\ \hline
$\Xi_c^0\to \Lambda^0K^{*0}$ & $\frac{1}{\sqrt6}(-a^\prime+2b^\prime-c^\prime+e^\prime-2f^\prime-2g^\prime)$ & $\Xi_c^0\to pK^{*-}$ & $(b^\prime+d^\prime-f^\prime)$\\ \hline
$\Xi_c^0\to\Sigma^0K^{*0}$ & $\frac{1}{\sqrt2}(a^\prime-c^\prime-e^\prime)$ & $\Xi_c^0\to n\overline{K}^{*0}$ & $(-a^\prime+b^\prime+e^\prime-f^\prime-g^\prime)$\\ \hline
$\Xi_c^0\to n\phi$ & $(a^\prime-e^\prime+g^\prime+h^\prime-r^\prime)$ & $\Xi_c^+\to\Sigma^+\omega$ & $\frac{\sqrt2}{2}(a^\prime+b^\prime+c^\prime+d^\prime+e^\prime+f^\prime+2h^\prime+2r^\prime)$\\ \hline
$\Xi_c^0\to p\rho^-$ & $(-b^\prime-d^\prime+f^\prime)$ & $\Xi_c^+\to\Lambda^0\rho^+$ & $\frac{1}{\sqrt6}(a^\prime+b^\prime+c^\prime+3d^\prime+e^\prime+f^\prime-2g^\prime)$\\ \hline
$\Xi_c^0\to n\rho^0$ & $\frac{1}{\sqrt2}(b^\prime-d^\prime-f^\prime)$ & $\Xi_c^+\to\Sigma^0\rho^+$ & $\frac{1}{\sqrt2}(a^\prime-b^\prime-c^\prime-d^\prime+e^\prime-f^\prime)$\\ \hline
$\Xi_c^0\to n\omega$ & $\frac{\sqrt2}{2}(-b^\prime-d^\prime+f^\prime-2h^\prime+2r^\prime)$ & $\Xi_c^+\to\Sigma^+\rho^0$ & $\frac{1}{\sqrt2}(-a^\prime+b^\prime-c^\prime-d^\prime-e^\prime+f^\prime)$\\ \hline
$\Xi_c^+\to p\omega$ & $\frac{\sqrt2}{2}(b^\prime+d^\prime+f^\prime+2h^\prime+2r^\prime)$  & $\Xi_c^+\to\Sigma^+\phi$ & $(c^\prime+g^\prime-h_1^\prime-r^\prime)$\\ \hline
$\Xi_c^+\to\Lambda^0K^{*+}$ & $\frac{1}{\sqrt6}(a^\prime-2b^\prime+c^\prime+e^\prime-2f^\prime-2g^\prime)$ & $\Xi_c^+\to p\overline{K}^{*0}$ & $(a^\prime-d^\prime+e^\prime-g^\prime)$\\ \hline
$\Xi_c^+\to\Sigma^0 K^{*+}$ & $\frac{1}{\sqrt2}(a^\prime-c^\prime+e^\prime)$ & $\Xi_c^+\to\Xi^0 K^{*+}$ & $(b^\prime-c^\prime+f^\prime+g^\prime)$\\ \hline
$\Xi_c^+\to\Sigma^+K^{*0}$ & $(a^\prime+c^\prime+e^\prime)$ &&
\end{tabular}
\end{ruledtabular}
\end{table*}

\begin{table*}[htp]
\caption{Decay amplitudes of $\mathcal{B}_c\rightarrow \mathcal{B}_{10}V$ modes in the $SU(3)_F$ limit.}\label{tab:amp4}
\begin{ruledtabular}
\footnotesize\begin{tabular}{cc|cc}
\textbf{Mode} & \textbf{Decay amplitude} & $\Xi_c^0\to\Sigma^{*0}K^{*0}$ & $\frac{1}{\sqrt6}(2\alpha^\prime-\beta^\prime+2\gamma^\prime-\delta^\prime)$ \\
\hline
$\Lambda_c^+\to\Sigma^{*+}\rho^0$ & $-\frac{1}{\sqrt6}(2\alpha^\prime-\beta^\prime+2\gamma^\prime+\delta^\prime)$ & $\Xi_c^0\to\Delta^+\rho^-$ & $\frac{1}{\sqrt3}(-\beta^\prime+2\gamma^\prime-\delta^\prime)$ \\ \hline
$\Lambda_c^+\to\Sigma^{*+}\phi$ & $\frac{1}{\sqrt3}(\beta^\prime-\delta^\prime-2\lambda^\prime)$ & $\Xi_c^0\to\Delta^-\rho^+$ & $(-\beta^\prime+\delta^\prime)$
\\ \hline
$\Lambda_c^+\to\Sigma^{*0}\rho^+$ & $-\frac{1}{\sqrt6}(2\alpha^\prime-\beta^\prime+2\gamma^\prime+\delta^\prime)$ & $\Xi_c^0\to\Sigma^{*-}K^{*+}$ & $\frac{1}{\sqrt3}(-\beta^\prime+\delta^\prime)$\\ \hline
$\Lambda_c^+\to\Delta^{++}K^{*-}$ & $(\beta^\prime+\delta^\prime)$ & $\Xi_c^0\to\Delta^0\rho^0$ & $\frac{2}{\sqrt6}(-\gamma^\prime+\delta^\prime)$ \\ \hline
$\Lambda_c^+\to \Delta^+\overline{K}^{*0}$ & $\frac{1}{\sqrt3}(\beta^\prime+\delta^\prime)$ & \textbf{Mode} & \textbf{Decay amplitude} (mod $s_1$)\\ \hline
$\Lambda_c^+\to\Xi^{*0} K^{*+}$ & $\frac{1}{\sqrt3}(\beta^\prime-2\gamma^\prime-\delta^\prime)$ & $\Lambda_c^+\to\Delta^+\rho^0$ & $\frac{2}{\sqrt6}(\alpha^\prime+\gamma^\prime+\delta^\prime)$\\ \hline
$\Lambda_c^+\to\Sigma^{*+}\omega$ & $\frac{1}{\sqrt6}(2\alpha^\prime+\beta^\prime-2\gamma^\prime-\delta^\prime+4\lambda^\prime)$ & $\Lambda_c^+\to\Delta^+\phi$ & $\frac{2}{3\sqrt3}(-2\alpha^\prime-2\beta^\prime+2\gamma^\prime-3\lambda^\prime)$\\ \hline
$\Xi_c^+\to\Sigma^{*+}\overline{K}^{*0}$ & $\frac{2}{\sqrt3}\alpha^\prime$ & $\Lambda_c^+\to\Delta^0\rho^+$ & $\frac{1}{\sqrt3}(2\alpha^\prime-\beta^\prime+2\gamma^\prime+\delta^\prime)$\\ \hline
$\Xi_c^+\to\Xi^{*0}\rho^+$ & $-\frac{2}{\sqrt3}\alpha^\prime$ & $\Lambda_c^+\to\Sigma^{*+}K^{*0}$ & $\frac{1}{\sqrt3}(2\alpha^\prime-\beta^\prime-\delta^\prime)$\\ \hline
$\Xi_c^0\to\Sigma^{*0}\overline{K}^{*0}$ & $\frac{1}{\sqrt6}(2\alpha^\prime-\beta^\prime+2\gamma^\prime-\delta^\prime)$ & $\Lambda_c^+\to \Sigma^{*0}K^{*+}$ & $\frac{1}{\sqrt6}(-2\alpha^\prime-\beta^\prime+2\gamma^\prime+\delta^\prime)$\\ \hline
$\Xi_c^0\to\Xi^{*0}\rho^0$ & $\frac{1}{\sqrt6}(2\alpha^\prime-\beta^\prime+\delta^\prime)$ & $\Lambda_c^+\to\Delta^{++} \rho^-$ & $(-\beta^\prime-\delta^\prime)$\\ \hline
$\Xi_c^0\to\Xi^{*0}\phi$ & $\frac{1}{\sqrt3}(\beta^\prime-2\gamma^\prime+\delta^\prime+2\lambda^\prime)$ & $\Lambda_c^+\to\Delta^+\omega$ & $\frac{\sqrt2}{3\sqrt3}(\alpha^\prime+\beta^\prime-\gamma^\prime+6\lambda^\prime)$\\ \hline
$\Xi_c^0\to\Sigma^{*+} K^{*-}$ & $\frac{1}{\sqrt3}(-\beta^\prime+2\gamma^\prime-\delta^\prime)$ & $\Xi_c^+\to\Sigma^{*+}\omega$ & $\frac{1}{\sqrt6}(\beta^\prime-2\gamma^\prime-\delta^\prime+4\lambda^\prime)$\\ \hline
$\Xi_c^0\to\Xi^{*-}\rho^+$ & $\frac{1}{\sqrt3}(-\beta^\prime+\delta^\prime)$ & $\Xi_c^+\to\Delta^+\overline{K}^{*0}$ & $\frac{1}{\sqrt3}(-2\alpha^\prime+\beta^\prime+\delta^\prime)$\\ \hline
$\Xi_c^0\to\Omega^-K^{*+}$ & $(-\beta^\prime+\delta^\prime)$ & $\Xi_c^+\to\Sigma^{*0}\rho^+$ & $\frac{1}{\sqrt6}(2\alpha^\prime+\beta^\prime-2\gamma^\prime-\delta^\prime)$\\ \hline
$\Xi_c^0\to\Xi^{*0}\omega$ & $\frac{1}{\sqrt6}(-2\alpha^\prime-\beta^\prime+\delta^\prime+4\lambda^\prime)$ & $\Xi_c^+\to\Sigma^{*+}\phi$ & $\frac{1}{\sqrt3}(-2\alpha^\prime-\beta^\prime-\delta^\prime-2\lambda^\prime)$\\ \hline
\textbf{Mode} & \textbf{Decay amplitude} (mod $s_1^2$) & $\Xi_c^+\to\Xi^{*0} K^{*+}$ & $\frac{1}{\sqrt3}(-2\alpha^\prime+\beta^\prime-2\gamma^\prime-\delta^\prime)$\\ \hline
$\Lambda_c^+\to\Delta^+K^{*0}$ & $-\frac{2}{\sqrt3}\alpha^\prime$ & $\Xi_c^+\to\Delta^{++}K^{*-}$ & $(\beta^\prime+\delta^\prime)$\\ \hline
$\Lambda_c^+\to\Delta^0 K^{*+}$ & $\frac{2}{\sqrt3}\alpha^\prime$ &  $\Xi_c^+\to\Sigma^{*+}\rho^0$ & $\frac{1}{\sqrt6}(\beta^\prime-2\gamma^\prime-\delta^\prime)$\\ \hline
$\Xi_c^+\to\Delta^+\phi$ & $\frac{2}{\sqrt3}(\alpha^\prime+\lambda^\prime)$ & $\Xi_c^0\to\Sigma^{*0}\omega$ & $\frac{1}{2\sqrt3}(2\alpha^\prime+3\beta^\prime-2\gamma^\prime-\delta^\prime+8\lambda^\prime)$\\ \hline
$\Xi_c^+\to\Sigma^{*0} K^{*+}$ & $\frac{1}{\sqrt6}(2\alpha^\prime-\beta^\prime+2\gamma^\prime+\delta^\prime)$ & $\Xi_c^0\to\Delta^0\overline{K}^{*0}$ & $\frac{1}{\sqrt3}(-2\alpha^\prime+\beta^\prime-2\gamma^\prime+\delta^\prime)$\\ \hline
$\Xi_c^+\to\Delta^{++}\rho^-$ & $(-\beta^\prime-\delta^\prime)$ & $\Xi_c^0\to\Delta^+K^{*-}$ & $\frac{1}{\sqrt3}(\beta^\prime-2\gamma^\prime+\delta^\prime)$\\ \hline
$\Xi_c^+\to\Sigma^{*+}K^{*0}$ & $\frac{1}{\sqrt3}(-\beta^\prime-\delta^\prime)$ & $\Xi_c^0\to\Sigma^{*0}\rho^0$ & $\frac{1}{2\sqrt3}(-2\alpha^\prime+\beta^\prime+2\gamma^\prime-3\delta^\prime)$\\ \hline
$\Xi_c^+\to\Delta^0\rho^+$ & $\frac{1}{\sqrt3}(-\beta^\prime+2\gamma^\prime+\delta^\prime)$ & $\Xi_c^0\to\Sigma^{*0}\phi$ & $\frac{1}{\sqrt6}(-2\alpha^\prime-\beta^\prime+2\gamma^\prime-\delta^\prime+4\lambda^\prime)$\\ \hline
$\Xi_c^+\to\Delta^+\rho^0$ & $\frac{2}{\sqrt6}(\gamma^\prime+\delta^\prime)$    & $\Xi_c^0\to\Xi^{*0} K^{*0}$ & $\frac{1}{\sqrt3}(-2\alpha^\prime+\beta^\prime-2\gamma^\prime+\delta^\prime)$\\  \hline
$\Xi_c^+\to\Delta^+\omega$ & $\sqrt{\frac{2}{3}}(-\beta^\prime+\gamma^\prime-2\lambda^\prime)$  & $\Xi_c^0\to\Sigma^{*+}\rho^-$ & $\frac{1}{\sqrt3}(\beta^\prime-2\gamma^\prime+\delta^\prime)$\\  \hline
$\Xi_c^0\to\Delta^0\omega$ & $\sqrt{\frac{2}{3}}(-\beta^\prime+\gamma^\prime-2\lambda^\prime)$  & $\Xi_c^0\to\Sigma^{*-}\rho^+$ & $\frac{2}{\sqrt3}(\beta^\prime-\delta^\prime)$\\  \hline
$\Xi_c^0\to\Delta^0\phi$ & $\frac{2}{\sqrt3}(\alpha^\prime+\lambda^\prime)$   &  $\Xi_c^0\to\Xi^{*-}K^{*+}$ & $\frac{2}{\sqrt3}(\beta^\prime-\delta^\prime)$
\end{tabular}
\end{ruledtabular}
\end{table*}

\clearpage
\section{Symmetry relations}\label{relation}

1. Isospin relations are listed following.\\
(1). Charmed baryon decays into one light meson and one octet baryon:
\begin{equation}
\mathcal{A}(\Xi_c^+\rightarrow\Xi^0\pi^+)+\mathcal{A}(\Xi_c^0\rightarrow\Xi^-\pi^+)+\sqrt{2}\mathcal{A}(\Xi_c^0\rightarrow\Xi^0\pi^0)=0,
\end{equation}
\begin{equation}
\mathcal{A}(\Xi_c^+\rightarrow\Sigma^+\overline{K}^0)+\mathcal{A}(\Xi_c^0\rightarrow\Sigma^+K^-)+\sqrt{2}\mathcal{A}(\Xi_c^0\rightarrow\Sigma^0\overline{K}^0)=0,
\end{equation}
\begin{align}
&\sqrt{2}\mathcal{A}(\Xi_c^+\rightarrow\Sigma^0\pi^+)+\sqrt{2}\mathcal{A}(\Xi_c^+\rightarrow\Sigma^+\pi^0)-\mathcal{A}(\Xi_c^0\rightarrow\Sigma^-\pi^+)
\nonumber\\&~~~~~~~~~~~~~~+{2}\mathcal{A}(\Xi_c^0\rightarrow\Sigma^0\pi^0)-\mathcal{A}(\Xi_c^0\rightarrow\Sigma^+\pi^-)=0,
\end{align}
\begin{equation}
\sqrt{2}\mathcal{A}(\Xi_c^+\rightarrow\Sigma^0K^+)-\mathcal{A}(\Xi_c^+\rightarrow\Sigma^+K^0)-\mathcal{A}(\Xi_c^0\rightarrow\Sigma^-K^+)
-\sqrt{2}\mathcal{A}(\Xi_c^0\rightarrow\Sigma^0K^0)=0,
\end{equation}
\begin{equation}
\sqrt{2}\mathcal{A}(\Xi_c^+\rightarrow p\pi^0)-\mathcal{A}(\Xi_c^+\rightarrow n\pi^+)-\mathcal{A}(\Xi_c^0\rightarrow p\pi^-)-
\sqrt{2}\mathcal{A}(\Xi_c^0\rightarrow n\pi^0)=0.
\end{equation}
(2). Charmed baryon decays into one light meson and one decuplet baryon:
\begin{equation}
\mathcal{A}(\Lambda_c^+\rightarrow\Sigma^{*+}\pi^0)-\mathcal{A}(\Lambda_c^+\rightarrow\Sigma^{*0}\pi^+)=0,
\end{equation}
\begin{equation}
\mathcal{A}(\Lambda_c^+\rightarrow\Delta^+K^0)+\mathcal{A}(\Lambda_c^+\rightarrow\Delta^0K^+)=0,
\end{equation}
\begin{equation}
\mathcal{A}(\Xi_c^+\to\Delta^+\eta_8)-\mathcal{A}(\Xi_c^0\to\Delta^0\eta_8)=0,
\end{equation}
\begin{equation}
\sqrt{6}\mathcal{A}(\Lambda^+_c\rightarrow \Delta^{+} \pi^0)-\sqrt{3}\mathcal{A}(\Lambda_c^+\rightarrow \Delta^0\pi^+) + \mathcal{A}(\Lambda_c^+\rightarrow\Delta^{++}\pi^-)=0,
\end{equation}
\begin{equation}
\sqrt{3}\mathcal{A}(\Xi_c^0 \rightarrow\Delta^+\pi^-) - \mathcal{A}(\Xi_c^0\rightarrow\Delta^-\pi^+) + \sqrt{6}\mathcal{A}(\Xi_c^0\rightarrow \Delta^0\pi^0)=0,
\end{equation}
\begin{equation}
\mathcal{A}(\Xi_c^+\rightarrow\Delta^{++}\pi^-) - \sqrt{3}\mathcal{A}(\Xi_c^+\rightarrow\Delta^0\pi^+)+ \sqrt{6}\mathcal{A}(\Xi_c^+\rightarrow\Delta^+\pi^0)=0,
\end{equation}
\begin{equation}
\mathcal{A}(\Xi_c^+\rightarrow\Sigma^{*+}\overline{K}^0)-\sqrt{2}\mathcal{A}(\Xi_c^0\rightarrow\Sigma^{*0}\overline{K}^0)
+\mathcal{A}(\Xi_c^0\rightarrow\Sigma^{*+}K^-)=0,
\end{equation}
\begin{equation}
\mathcal{A}(\Xi_c^+\rightarrow\Xi^{*0}\pi^+)+\sqrt{2}\mathcal{A}(\Xi_c^0\rightarrow\Xi^{*0}\pi^0)-\mathcal{A}(\Xi_c^0\rightarrow\Xi^{*-}\pi^+)=0,
\end{equation}
\begin{equation}
\sqrt{3}\mathcal{A}(\Xi_c^+\rightarrow\Delta^+\overline{K}^0)-\mathcal{A}(\Xi_c^+\rightarrow\Delta^{++}K^-)-
\sqrt{3}\mathcal{A}(\Xi_c^0\rightarrow\Delta^0\overline{K}^0)+\sqrt{3 }\mathcal{A}(\Xi_c^0\rightarrow\Delta^+K^-)=0,
\end{equation}
\begin{equation}
\sqrt{2}\mathcal{A}(\Xi_c^+\rightarrow\Sigma^{*0}K^+)+\mathcal{A}(\Xi_c^+\rightarrow\Sigma^{*+}K^0)-\sqrt{2}\mathcal{A}(\Xi_c^0\rightarrow\Sigma^{*0}K^0)
-\mathcal{A}(\Xi_c^0\rightarrow\Sigma^{*-}K^+)=0,
\end{equation}
\begin{equation}
\sqrt{2}\mathcal{A}(\Xi_c^+\rightarrow\Delta^{++}\pi^-)
+\sqrt{3}\mathcal{A}(\Xi_c^+\rightarrow\Delta^+\pi^0)
-\sqrt{2}\mathcal{A}(\Xi_c^0\rightarrow\Delta^-\pi^+)
+\sqrt{3}\mathcal{A}(\Xi_c^0\rightarrow\Delta^0\pi^0)=0,
\end{equation}
\begin{equation}
\mathcal{A}(\Xi_c^+\rightarrow\Delta^{++}\pi^-)+\sqrt{3}\mathcal{A}(\Xi_c^+\rightarrow\Delta^0\pi^+)-\sqrt{3}\mathcal{A}(\Xi_c^0\rightarrow\Delta^+\pi^-)-\mathcal{A}(\Xi_c^0\rightarrow\Delta^-\pi^+)=0,
\end{equation}
\begin{equation}
\sqrt{2}\mathcal{A}(\Xi_c^+\rightarrow\Delta^0\pi^+)-\mathcal{A}(\Xi_c^+\rightarrow\Delta^+\pi^0)-\sqrt{2}\mathcal{A}(\Xi_c^0\rightarrow\Delta^+\pi^-)-\mathcal{A}(\Xi_c^0\rightarrow\Delta^0\pi^0)=0,
\end{equation}
\begin{align}
&\sqrt{2}\mathcal{A}(\Xi_c^+\rightarrow\Sigma^{*0}\pi^+)-\sqrt{2}\mathcal{A}(\Xi_c^+\rightarrow\Sigma^{*+}\pi^0)-\mathcal{A}(\Xi_c^0\rightarrow\Sigma^{*-}\pi^+)
+{2}\mathcal{A}(\Xi_c^0\rightarrow\Sigma^{*0}\pi^0)
\nonumber\\&~~~~~~~~~~~~~~~~~~~~~~~~~~+\mathcal{A}(\Xi_c^0\rightarrow\Sigma^{*+}\pi^-)=0.
\end{align}

2. $U$-spin relations are listed following.\\
(1). Charmed baryon decays into one light meson and one octet baryon:
\begin{equation}
\mathcal{A}(\Xi_c^0\rightarrow\Xi^0K^0)
+\mathcal{A}(\Xi_c^0\rightarrow n\overline{K}^0)=0,
\end{equation}
\begin{equation}
\mathcal{A}(\Lambda_c^+\rightarrow n\pi^+)
-\mathcal{A}(\Xi_c^+\rightarrow\Xi^0K^+)=0,
\end{equation}
\begin{equation}
\mathcal{A}(\Lambda_c^+\rightarrow nK^+)
-\sin^2\theta\mathcal{A}(\Xi_c^+\rightarrow\Xi^0\pi^+)=0,
\end{equation}
\begin{equation}
\mathcal{A}(\Lambda_c^+\rightarrow pK^0)
-\sin^2\theta\mathcal{A}(\Xi_c^+\rightarrow\Sigma^+\overline{K}^0)=0,
\end{equation}
\begin{equation}
\mathcal{A}(\Lambda_c^+\rightarrow \Sigma^+K^0)
-\mathcal{A}(\Xi_c^+\rightarrow p\overline{K}^0)=0,
\end{equation}
\begin{equation}
\sin^2\theta\mathcal{A}(\Xi_c^0 \rightarrow \Xi^- \pi^+)
=-\sin\theta\mathcal{A}(\Xi_c^0 \rightarrow \Sigma^-\pi^+)
=\sin\theta\mathcal{A}(\Xi_c^0 \rightarrow \Xi^- K^+)
=-\mathcal{A}(\Xi_c^0 \rightarrow \Sigma^- K^+),
\end{equation}
\begin{equation}
\sin^2\theta\mathcal{A}(\Xi_c^0\rightarrow\Sigma^+K^-)
=-\sin\theta\mathcal{A}(\Xi_c^0\rightarrow\Sigma^+\pi^-)
=\sin\theta\mathcal{A}(\Xi_c^0\rightarrow pK^-)
=-\mathcal{A}(\Xi_c^0\rightarrow p\pi^-),
\end{equation}
\begin{equation}
\sin\theta\mathcal{A}(\Lambda^+_c\rightarrow n\pi^+)
-\mathcal{A}(\Lambda^+_c\rightarrow n K^+)
-\sin^2\theta\mathcal{A}(\Lambda^+_c\rightarrow\Xi^0K^+)=0,
\end{equation}
\begin{equation}
\sin^2\theta\mathcal{A}(\Lambda_c^+\rightarrow p\overline{K}^0)
+\mathcal{A}(\Lambda_c^+\rightarrow pK^0)
-\sin\theta\mathcal{A}(\Lambda_c^+\rightarrow\Sigma^+K^0)=0,
\end{equation}
\begin{equation}
\sqrt2\sin^2\theta\mathcal{A}(\Xi_c^0\rightarrow\Xi^0\pi^0)
+\sin\theta\mathcal{A}(\Xi_c^0\rightarrow\Xi^0K^0)
+\sqrt2\mathcal{A}(\Xi_c^0\rightarrow n\pi^0)=0,
\end{equation}
\begin{equation}
\sqrt2\sin^2\theta\mathcal{A}(\Xi_c^0\rightarrow\Xi^0\pi^0)
-\sin\theta\mathcal{A}(\Xi_c^0\rightarrow n\overline{K}^0)
+\sqrt2\mathcal{A}(\Xi_c^0\rightarrow n\pi^0)=0,
\end{equation}
\begin{equation}
\sin\theta\mathcal{A}(\Xi_c^0\rightarrow\Xi^0\pi^0)
+\sin\theta\mathcal{A}(\Xi_c^0\rightarrow\Sigma^0\overline{K}^0)
-\sqrt2\mathcal{A}(\Xi_c^0\rightarrow\Sigma^0\pi^0)=0,
\end{equation}
\begin{equation}
\sin\theta\mathcal{A}(\Xi_c^0\rightarrow\Xi^0K^0)
-\sqrt2\sin^2\theta\mathcal{A}(\Xi_c^0\rightarrow\Sigma^0\overline{K}^0)
-\sqrt2\mathcal{A}(\Xi_c^0\rightarrow\Sigma^0K^0)=0,
\end{equation}
\begin{equation}
\sqrt2\sin^2\theta\mathcal{A}(\Xi_c^0\rightarrow\Sigma^0\overline{K}^0)
+\sqrt2\mathcal{A}(\Xi_c^0\rightarrow\Sigma^0K^0)
+\sin\theta\mathcal{A}(\Xi_c^0\rightarrow n\overline{K}^0)=0,
\end{equation}
\begin{equation}
\sqrt2\sin\theta\mathcal{A}(\Xi_c^0\rightarrow\Sigma^0\pi^0)
+\mathcal{A}(\Xi_c^0\rightarrow\Sigma^0K^0)
+\mathcal{A}(\Xi_c^0\rightarrow n\pi^0)=0,
\end{equation}
\begin{equation}
\sin^2\theta\mathcal{A}(\Xi_c^+\rightarrow\Xi^0\pi^+)
-\sin\theta\mathcal{A}(\Xi_c^+\rightarrow\Xi^0K^+)
+\mathcal{A}(\Xi_c^+\rightarrow n\pi^+)=0,
\end{equation}
\begin{equation}
\sqrt2\sin\theta\mathcal{A}(\Xi_c^+\rightarrow\Sigma^0\pi^+)
-\sqrt2\mathcal{A}(\Xi_c^+\rightarrow\Sigma^0K^+)
+\mathcal{A}(\Xi_c^+\rightarrow n\pi^+)=0,
\end{equation}
\begin{equation}
\sin^2\theta\mathcal{A}(\Xi_c^+\rightarrow\Sigma^+\overline{K}^0)
+\mathcal{A}(\Xi_c^+\rightarrow\Sigma^+K^0)
-\sin\theta\mathcal{A}(\Xi_c^+\rightarrow p\overline{K}^0)=0,
\end{equation}
\begin{equation}
\sqrt2\sin\theta\mathcal{A}(\Xi_c^+\rightarrow\Sigma^+\pi^0)
+\mathcal{A}(\Xi_c^+\rightarrow\Sigma^+K^0)
-\sqrt2\mathcal{A}(\Xi_c^+\rightarrow p\pi^0)=0,
\end{equation}
\begin{equation}
\sqrt2\sin^2\theta\mathcal{A}(\Lambda_c^+\rightarrow\Sigma^0\pi^+)
-\sqrt2\sin\theta\mathcal{A}(\Lambda_c^+\rightarrow\Sigma^0K^+)
+\mathcal{A}(\Xi_c^+\rightarrow n\pi^+)=0,
\end{equation}
\begin{equation}
\sqrt2\sin\theta\mathcal{A}(\Lambda_c^+\rightarrow\Sigma^0K^+)
+\mathcal{A}(\Lambda_c^+\rightarrow nK^+)
-\sqrt2\mathcal{A}(\Xi_c^+\rightarrow\Sigma^0K^+)=0,
\end{equation}
\begin{equation}
\mathcal{A}(\Lambda_c^+\rightarrow nK^+)
+\sin^2\theta\mathcal{A}(\Lambda_c^+\rightarrow\Xi^0K^+)
-\sin\theta\mathcal{A}(\Xi_c^+\rightarrow\Xi^0K^+)=0,
\end{equation}
\begin{equation}
\sin^2\theta\mathcal{A}(\Lambda_c^+\rightarrow\Xi^0K^+)
+\sqrt2\sin\theta\mathcal{A}(\Xi_c^+\rightarrow\Sigma^0\pi^+)
-\sqrt2\mathcal{A}(\Xi_c^+\rightarrow\Sigma^0K^+)=0,
\end{equation}
\begin{equation}
\mathcal{A}(\Lambda_c^+\rightarrow nK^+)
-\sin\theta\mathcal{A}(\Xi_c^+\rightarrow\Xi^0K^+)
+\mathcal{A}(\Xi_c^+\rightarrow n\pi^+)=0,
\end{equation}
\begin{equation}
\sqrt2\sin^2\theta\mathcal{A}(\Lambda_c^+\rightarrow\Sigma^0\pi^+)
+\sin\theta\mathcal{A}(\Lambda_c^+\rightarrow n\pi^+)
-\sqrt2\mathcal{A}(\Xi_c^+\rightarrow\Sigma^0K^+)=0,
\end{equation}
\begin{equation}
\sin^2\theta\mathcal{A}(\Lambda_c^+\rightarrow p\overline{K}^0)
+\mathcal{A}(\Lambda_c^+\rightarrow pK^0)
-\sin\theta\mathcal{A}(\Xi_c^+\rightarrow p\overline{K}^0)=0,
\end{equation}
\begin{equation}
\sin\theta\mathcal{A}(\Lambda_c^+\rightarrow p\overline{K}^0)
-\mathcal{A}(\Lambda_c^+\rightarrow\Sigma^+K^0)
+\sin\theta\mathcal{A}(\Xi_c^+\rightarrow\Sigma^+\overline{K}^0)=0,
\end{equation}
\begin{equation}
\sin^2\theta\mathcal{A}(\Lambda_c^+\rightarrow p\overline{K}^0)
+\sqrt2\sin\theta\mathcal{A}(\Xi_c^+\rightarrow\Sigma^+\pi^0)
-\sqrt2\mathcal{A}(\Xi_c^+\rightarrow p\pi^0)=0,
\end{equation}
\begin{equation}
\sqrt2\sin\theta\mathcal{A}(\Lambda_c^+\rightarrow p\pi^0)
+\mathcal{A}(\Lambda_c^+\rightarrow pK^0)
-\sqrt2\mathcal{A}(\Xi_c^+\rightarrow p\pi^0)=0,
\end{equation}
\begin{equation}
\sqrt2\sin\theta\mathcal{A}(\Lambda_c^+\rightarrow p\pi^0)
-\sqrt2\sin^2\theta\mathcal{A}(\Lambda_c^+\rightarrow\Sigma^+\pi^0)
-\mathcal{A}(\Xi_c^+\rightarrow\Sigma^+K^0)=0,
\end{equation}
\begin{equation}
\mathcal{A}(\Lambda_c^+\rightarrow pK^0)
+\sqrt2\sin^2\theta\mathcal{A}(\Lambda_c^+\rightarrow\Sigma^+\pi^0)
-\sqrt2\sin\theta\mathcal{A}(\Xi_c^+\rightarrow\Sigma^+\pi^0)=0,
\end{equation}
\begin{equation}
\sqrt2\sin^2\theta\mathcal{A}(\Lambda_c^+\rightarrow\Sigma^+\pi^0)
+\sin\theta\mathcal{A}(\Lambda_c^+\rightarrow\Sigma^+K^0)
-\sqrt2\mathcal{A}(\Xi_c^+\rightarrow p\pi^0)=0,
\end{equation}
\begin{equation}
\sin^2\theta\mathcal{A}(\Lambda_c^+\rightarrow\Lambda^0\pi^+)
-\sin\theta\mathcal{A}(\Lambda_c^+\rightarrow\Lambda^0K^+)
-\sin\theta\mathcal{A}(\Xi_c^+\rightarrow\Lambda^0\pi^+)
+\mathcal{A}(\Xi_c^+\rightarrow\Lambda^0K^+)=0,
\end{equation}
\begin{equation}
\sin\theta\mathcal{A}(\Lambda_c^+\rightarrow p\eta_8)
-\sin^2\theta\mathcal{A}(\Lambda_c^+\rightarrow\Sigma^+\eta_8)
+\sin\theta\mathcal{A}(\Xi_c^+\rightarrow\Sigma^+\eta_8)
-\mathcal{A}(\Xi_c^+\rightarrow p\eta_8)=0.
\end{equation}
(2). Charmed baryon decays into one light meson and one decuplet baryon:
\begin{equation}
\sqrt2\sin\theta\mathcal{A}(\Lambda_c^+\rightarrow\Sigma^{*0}\pi^+)
-\mathcal{A}(\Lambda_c^+\rightarrow\Delta^0\pi^+)=0,
\end{equation}
\begin{equation}
\sin\theta\mathcal{A}(\Xi_c^+\rightarrow\Xi^{*0}K^+)
+\sqrt2\mathcal{A}(\Xi_c^+\rightarrow\Sigma^{*0}K^+)=0,
\end{equation}
\begin{equation}
\sin\theta\mathcal{A}(\Xi_c^+\rightarrow\Delta^{++}K^-)
-\mathcal{A}(\Xi_c^+\rightarrow\Delta^{++}\pi^-)=0,
\end{equation}
\begin{equation}
\mathcal{A}(\Lambda_c^+\rightarrow\Sigma^{*+}K^0)
+\mathcal{A}(\Xi_c^+\rightarrow\Delta^+\overline{K}^0)=0,
\end{equation}
\begin{equation}
\mathcal{A}(\Lambda_c^+\rightarrow\Delta^+K^0)
+\sin^2\theta\mathcal{A}(\Xi_c^+\rightarrow\Sigma^{*+}\overline{K}^0)=0,
\end{equation}
\begin{equation}
\sqrt2\sin\theta\mathcal{A}(\Lambda_c^+\rightarrow\Sigma^{*0}\pi^+)
+\mathcal{A}(\Xi_c^+\rightarrow\Xi^{*0}K^+)=0,
\end{equation}
\begin{equation}
\mathcal{A}(\Lambda_c^+\rightarrow\Sigma^{*0}K^+)
+\mathcal{A}(\Xi_c^+\rightarrow\Sigma^{*0}\pi^+)=0,
\end{equation}
\begin{equation}
\sin\theta\mathcal{A}(\Lambda_c^+\rightarrow\Delta^0\pi^+)
+\sqrt2\mathcal{A}(\Xi_c^+\rightarrow\Sigma^{*0}K^+)=0,
\end{equation}
\begin{equation}
\mathcal{A}(\Lambda_c^+\rightarrow\Delta^0K^+)
+\sin^2\theta\mathcal{A}(\Xi_c^+\rightarrow\Xi^{*0}\pi^+)=0,
\end{equation}
\begin{equation}
\sqrt2\sin^2\theta\mathcal{A}(\Xi_c^0\rightarrow\Sigma^{*0}\overline{K}^0)
=\sin\theta\mathcal{A}(\Xi_c^0\rightarrow\Delta^0\overline{K}^0)
=\sin\theta\mathcal{A}(\Xi_c^0\rightarrow\Xi^{*0}K^0)
=\sqrt2\mathcal{A}(\Xi_c^0\rightarrow\Sigma^{*0}K^0),
\end{equation}
\begin{equation}
\begin{split}
&\sin^2\theta\mathcal{A}(\Xi_c^0\rightarrow\Sigma^{*+}K^-)
=\sin\theta\mathcal{A}(\Xi_c^0\rightarrow\Sigma^{*+}\pi^-)
=\sin\theta\mathcal{A}(\Xi_c^0\rightarrow\Delta^+K^-)
=\mathcal{A}(\Xi_c^0\rightarrow\Delta^+\pi^-),
\end{split}
\end{equation}
\begin{align}
&~~~~2\sqrt3\sin^2\theta\mathcal{A}(\Xi_c^0\rightarrow\Xi^{*-}\pi^+)
=\sqrt3\sin\theta\mathcal{A}(\Xi_c^0\rightarrow\Xi^{*-}K^+)
=2\mathcal{A}(\Xi_c^0\rightarrow\Delta^-\pi^+)
\nonumber\\&=\sqrt3\sin\theta\mathcal{A}(\Xi_c^0\rightarrow\Sigma^{*-}\pi^+)
=2\sqrt3\mathcal{A}(\Xi_c^0\rightarrow\Sigma^{*-}K^+)
=2\sin^2\theta\mathcal{A}(\Xi_c^0\rightarrow\Omega^-K^+),
\end{align}
\begin{equation}
\sqrt2\sin\theta\mathcal{A}(\Lambda_c^+\rightarrow\Sigma^{*+}\pi^0)
-\sqrt2\mathcal{A}(\Lambda_c^+\rightarrow\Delta^+\pi^0)
-\sin\theta\mathcal{A}(\Lambda_c^+\rightarrow\Delta^+\overline{K}^0)=0,
\end{equation}
\begin{equation}
\mathcal{A}(\Lambda_c^+\rightarrow\Sigma^{*+}K^0)
-\mathcal{A}(\Lambda_c^+\rightarrow\Delta^+K^0)
-\sin^2\theta\mathcal{A}(\Lambda_c^+\rightarrow\Delta^+\overline{K}^0)=0,
\end{equation}
\begin{equation}
\sin^2\theta\mathcal{A}(\Lambda_c^+\rightarrow\Sigma^{*0}\pi^+)
-\sin\theta\mathcal{A}(\Lambda_c^+\rightarrow\Sigma^{*0}K^+)
+\sqrt2\mathcal{A}(\Lambda_c^+\rightarrow\Delta^0K^+)=0,
\end{equation}
\begin{equation}
\sin\theta\mathcal{A}(\Lambda_c^+\rightarrow\Sigma^{*0}\pi^+)
+\mathcal{A}(\Lambda_c^+\rightarrow\Sigma^{*0}K^+)
-\sqrt2\sin\theta\mathcal{A}(\Lambda_c^+\rightarrow\Xi^{*0}K^+)=0,
\end{equation}
\begin{equation}
\sqrt2\sin^2\theta\mathcal{A}(\Lambda_c^+\rightarrow\Sigma^{*0}\pi^+)
-\sin\theta\mathcal{A}(\Lambda_c^+\rightarrow\Delta^0\pi^+)
-2\mathcal{A}(\Lambda_c^+\rightarrow\Delta^0K^+)=0,
\end{equation}
\begin{equation}
\sqrt2\sin^2\theta\mathcal{A}(\Lambda_c^+\rightarrow\Sigma^{*0}\pi^+)
+\mathcal{A}(\Lambda_c^+\rightarrow\Delta^0K^+)
-\sin^2\theta\mathcal{A}(\Lambda_c^+\rightarrow\Xi^{*0}K^+)=0,
\end{equation}
\begin{equation}
\sqrt2\sin\theta\mathcal{A}(\Lambda_c^+\rightarrow\Sigma^{*0}K^+)
-\sin\theta\mathcal{A}(\Lambda_c^+\rightarrow\Delta^0\pi^+)
-2\mathcal{A}(\Lambda_c^+\rightarrow\Delta^0K^+)=0,
\end{equation}
\begin{equation}
\sqrt2\mathcal{A}(\Lambda_c^+\rightarrow\Sigma^{*0}K^+)
+\mathcal{A}(\Lambda_c^+\rightarrow\Delta^0\pi^+)
-2\sin\theta\mathcal{A}(\Lambda_c^+\rightarrow\Xi^{*0}K^+)=0,
\end{equation}
\begin{equation}
\sqrt2\sin\theta\mathcal{A}(\Lambda_c^+\rightarrow\Sigma^{*0}K^+)
-\mathcal{A}(\Lambda_c^+\rightarrow\Delta^0K^+)
-\sin^2\theta\mathcal{A}(\Lambda_c^+\rightarrow\Xi^{*0}K^+)=0,
\end{equation}
\begin{equation}
\sin^2\theta\mathcal{A}(\Xi_c^0\rightarrow\Xi^{*0}\pi^0)
-\sin\theta\mathcal{A}(\Xi_c^0\rightarrow\Sigma^{*0}\pi^0)
+\mathcal{A}(\Xi_c^0\rightarrow\Delta^0\pi^0)=0,
\end{equation}
\begin{equation}
\sin^2\theta\mathcal{A}(\Xi_c^0\rightarrow\Sigma^{*0}\overline{K}^0)
-\sin^2\theta\mathcal{A}(\Xi_c^0\rightarrow\Xi^{*0}\pi^0)
+\mathcal{A}(\Xi_c^0\rightarrow\Delta^0\pi^0)=0,
\end{equation}
\begin{equation}
\sin^2\theta\mathcal{A}(\Xi_c^0\rightarrow\Sigma^{*0}\overline{K}^0)
-\sqrt2\sin\theta\mathcal{A}(\Xi_c^0\rightarrow\Sigma^{*0}\pi^0)
+2\mathcal{A}(\Xi_c^0\rightarrow\Delta^0\pi^0)=0,
\end{equation}
\begin{equation}
\sin^2\theta\mathcal{A}(\Xi_c^0\rightarrow\Xi^{*0}\eta_8)
-\sqrt2\sin\theta\mathcal{A}(\Xi_c^0\rightarrow\Sigma^{*0}\eta_8)
+\mathcal{A}(\Xi_c^0\rightarrow\Delta^0\eta_8)=0,
\end{equation}
\begin{equation}
\sin^2\theta\mathcal{A}(\Xi_c^+\rightarrow\Sigma^{*+}\overline{K}^0)
+\mathcal{A}(\Xi_c^+\rightarrow\Sigma^{*+}K^0)
-\sin\theta\mathcal{A}(\Xi_c^+\rightarrow\Delta^+\overline{K}^0)=0,
\end{equation}
\begin{equation}
\sqrt2\sin\theta\mathcal{A}(\Xi_c^+\rightarrow\Sigma^{*+}\pi^0)
-\mathcal{A}(\Xi_c^+\rightarrow\Sigma^{*+}K^0)
-\sqrt2\mathcal{A}(\Xi_c^+\rightarrow\Delta^+\pi^0)=0,
\end{equation}
\begin{equation}
\sin^2\theta\mathcal{A}(\Xi_c^+\rightarrow\Xi^{*0}\pi^+)
+\sin\theta\mathcal{A}(\Xi_c^+\rightarrow\Xi^{*0}K^+)
-\mathcal{A}(\Xi_c^+\rightarrow\Delta^0\pi^+)=0,
\end{equation}
\begin{equation}
\sin^2\theta\mathcal{A}(\Xi_c^+\rightarrow\Xi^{*0}\pi^+)
-\sin\theta\mathcal{A}(\Xi_c^+\rightarrow\Sigma^{*0}\pi^+)
+\mathcal{A}(\Xi_c^+\rightarrow\Delta^0\pi^+)=0,
\end{equation}
\begin{equation}
\sin^2\theta\mathcal{A}(\Xi_c^+\rightarrow\Xi^{*0}\pi^+)
+\sqrt2\mathcal{A}(\Xi_c^+\rightarrow\Sigma^{*0}K^+)
-\mathcal{A}(\Xi_c^+\rightarrow\Delta^0\pi^+)=0,
\end{equation}
\begin{equation}
\sin\theta\mathcal{A}(\Xi_c^+\rightarrow\Xi^{*0}K^+)
+\sqrt2\sin\theta\mathcal{A}(\Xi_c^+\rightarrow\Sigma^{*0}\pi^+)
-2\mathcal{A}(\Xi_c^+\rightarrow\Delta^0\pi^+)=0,
\end{equation}
\begin{equation}
\sin\theta\mathcal{A}(\Xi_c^+\rightarrow\Sigma^{*0}\pi^+)
+\mathcal{A}(\Xi_c^+\rightarrow\Sigma^{*0}K^+)
-\sqrt2\mathcal{A}(\Xi_c^+\rightarrow\Delta^0\pi^+)=0,
\end{equation}
\begin{equation}
2\sin\theta\mathcal{A}(\Xi_c^+\rightarrow\Xi^{*0}\pi^+)
+\mathcal{A}(\Xi_c^+\rightarrow\Xi^{*0}K^+)
-\sqrt2\mathcal{A}(\Xi_c^+\rightarrow\Sigma^{*0}\pi^+)=0,
\end{equation}
\begin{equation}
\sqrt2\sin^2\theta\mathcal{A}(\Lambda_c^+\rightarrow\Sigma^{*+}\pi^0)
-\sin\theta\mathcal{A}(\Lambda_c^+\rightarrow\Sigma^{*+}K^0)
+\sqrt2\mathcal{A}(\Xi_c^+\rightarrow\Delta^+\pi^0)=0,
\end{equation}
\begin{equation}
\sqrt2\sin^2\theta\mathcal{A}(\Lambda_c^+\rightarrow\Sigma^{*+}\pi^0)
-\sqrt2\sin\theta\mathcal{A}(\Lambda_c^+\rightarrow\Delta^+\pi^0)
+\mathcal{A}(\Xi_c^+\rightarrow\Sigma^{*+}K^0)=0,
\end{equation}
\begin{equation}
\sqrt2\sin^2\theta\mathcal{A}(\Lambda_c^+\rightarrow\Sigma^{*+}\pi^0)
-\mathcal{A}(\Lambda_c^+\rightarrow\Delta^+K^0)
+\sqrt2\sin\theta\mathcal{A}(\Xi_c^+\rightarrow\Sigma^{*+}\pi^0)=0,
\end{equation}
\begin{equation}
\sin\theta\mathcal{A}(\Lambda_c^+\rightarrow\Sigma^{*+}K^0)
-\mathcal{A}(\Lambda_c^+\rightarrow\Delta^+K^0)
+\mathcal{A}(\Xi_c^+\rightarrow\Sigma^{*+}K^0)=0,
\end{equation}
\begin{equation}
\sqrt2\sin\theta\mathcal{A}(\Lambda_c^+\rightarrow\Delta^+\pi^0)
-\mathcal{A}(\Lambda_c^+\rightarrow\Delta^+K^0)
+\sqrt2\mathcal{A}(\Xi_c^+\rightarrow\Delta^+\pi^0)=0,
\end{equation}
\begin{equation}
\mathcal{A}(\Lambda_c^+\rightarrow\Delta^+K^0)
+\sin^2\theta\mathcal{A}(\Lambda_c^+\rightarrow\Delta^+\overline{K}^0)
+\sin\theta\mathcal{A}(\Xi_c^+\rightarrow\Delta^+\overline{K}^0)=0,
\end{equation}
\begin{equation}
\sin^2\theta\mathcal{A}(\Lambda_c^+\rightarrow\Delta^+\overline{K}^0)
+\sqrt2\sin\theta\mathcal{A}(\Xi_c^+\rightarrow\Sigma^{*+}\pi^0)
-\sqrt2\mathcal{A}(\Xi_c^+\rightarrow\Delta^+\pi^0)=0.
\end{equation}

3. $V$-spin relations are listed following.\\
(1). Charmed baryon decays into one light meson and one octet baryon:
\begin{equation}
\mathcal{A}(\Xi_c^+\to p\pi^0)+\sqrt3\mathcal{A}(\Xi_c^+\to p\eta_8)+\mathcal{A}(\Xi_c^+\to\Sigma^0 K^+)+\sqrt3\mathcal{A}(\Xi_c^+\to\Lambda^0 K^+)=0,
\end{equation}
\begin{equation}
\sqrt2\mathcal{A}(\Lambda_c^+\to nK^+)-\sqrt2\mathcal{A}(\Xi_c^0\to\Sigma^-K^+)-\mathcal{A}(\Xi_c^0\to n\pi^0)-\sqrt3\mathcal{A}(\Xi_c^0\to n\eta_8)=0,
\end{equation}
\begin{equation}
2\sqrt2\mathcal{A}(\Lambda_c^+\to pK^0)-2\sqrt2\mathcal{A}(\Xi_c^0\to p\pi^-)-\mathcal{A}(\Xi_c^0\to\Sigma^0K^0)-\sqrt3\mathcal{A}(\Xi_c^0\to\Lambda^0K^0)=0,
\end{equation}
\begin{align}
&\sqrt2\mathcal{A}(\Lambda_c^+\to\Sigma^0K^+)+\sqrt6\mathcal{A}(\Lambda_c^+\to\Lambda^0K^+)+\sqrt2\mathcal{A}(\Lambda_c^+\to p\pi^0)+\sqrt6\mathcal{A}(\Lambda_c^+\to p\eta_8)\nonumber\\
&+2\mathcal{A}(\Xi_c^0\to\Xi^-K^+)-\mathcal{A}(\Xi_c^0\to\Sigma^0\pi^0)
-\sqrt3\mathcal{A}(\Xi_c^0\to\Sigma^0\eta_8)
-\sqrt3\mathcal{A}(\Xi_c^0\to\Lambda^0\pi^0)\nonumber\\
&~~~~~~~~~~~~~~~~~~~~~~~~~~~+3\mathcal{A}(\Xi_c^0\to\Lambda^0\eta_8)+2\mathcal{A}(\Xi_c^0\to pK^-)=0,
\end{align}
\begin{align}
&\sqrt2\mathcal{A}(\Lambda_c^+\to\Sigma^0\pi^+)+\sqrt6\mathcal{A}(\Lambda_c^+\to\Lambda^0\pi^+)-2\mathcal{A}(\Lambda_c^+\to p\overline{K}^0)+2\mathcal{A}(\Xi_c^0\to\Xi^-\pi^+)\nonumber\\
&~~~~~~~~~~~~~~~~~~~~~~~~~~~+\sqrt2\mathcal{A}(\Xi_c^0\to\Sigma^0\overline{K}^0)+\sqrt6\mathcal{A}(\Xi_c^0\to\Lambda^0\overline{K}^0)=0,
\end{align}
\begin{align}
&\sqrt2\mathcal{A}(\Lambda_c^+\to\Sigma^+\pi^0)+\sqrt6\mathcal{A}(\Lambda_c^+\to\Sigma^+\eta_8)
-2\mathcal{A}(\Lambda_c^+\to\Xi^0K^+)+2\mathcal{A}(\Xi_c^0\to\Sigma^+K^-)\nonumber\\
&~~~~~~~~~~~~~~~~~~~~~~~~~~~+\sqrt2\mathcal{A}(\Xi_c^0\to\Xi^0\pi^0)+\sqrt6\mathcal{A}(\Xi_c^0\to\Xi^0\eta_8)=0.
\end{align}
(2). Charmed baryon decays into one light meson and one decuplet baryon:
\begin{equation}
\mathcal{A}(\Xi_c^0\rightarrow\Sigma^{*0}\pi^0)-
\sqrt{3}\mathcal{A}(\Xi_c^0\rightarrow\Sigma^{*0}\eta_8)=0,
\end{equation}
\begin{equation}
\mathcal{A}(\Xi_c^+\rightarrow\Delta^{++}\pi^-)
-\sqrt{3}\mathcal{A}(\Xi_c^+\rightarrow\Sigma^{*+}K^0)=0.
\end{equation}
\begin{equation}
\mathcal{A}(\Lambda_c^+\rightarrow\Delta^+ K^0)
+\sqrt{2}\mathcal{A}(\Xi_c^0\rightarrow\Sigma^{*0}K^0)
-\mathcal{A}(\Xi_c^0\rightarrow\Delta^+\pi^-)=0,
\end{equation}
\begin{equation}
2\sqrt6\mathcal{A}(\Lambda_c^+\rightarrow\Delta^+\eta_8)
-4\mathcal{A}(\Xi_c^0\rightarrow\Sigma^{*0}\pi^0)
+3\mathcal{A}(\Xi_c^0\rightarrow\Xi^{*-}K^+)=0,
\end{equation}
\begin{equation}
2\mathcal{A}(\Xi_c^+\rightarrow\Sigma^{*0}K^+)-
\mathcal{A}(\Xi_c^+\rightarrow\Delta^{+}\pi^0)-
\sqrt{3}\mathcal{A}(\Xi_c^+\rightarrow\Delta^+\eta_8)=0,
\end{equation}
\begin{equation}
\mathcal{A}(\Lambda_c^+\rightarrow\Delta^{++}K^-)+
\sqrt3\mathcal{A}(\Lambda_c^+\rightarrow\Xi^{*0}K^+)+
\sqrt3\mathcal{A}(\Xi_c^0\rightarrow\Sigma^{*+}K^-)+
\mathcal{A}(\Xi_c^0\rightarrow\Omega^-K^+)=0,
\end{equation}
\begin{equation}
\sqrt6\mathcal{A}(\Lambda_c^+\rightarrow\Sigma^{*+}\pi^0)+
\sqrt3\mathcal{A}(\Lambda_c^+\rightarrow\Xi^{*0}K^+)+
\sqrt6\mathcal{A}(\Xi_c^0\rightarrow \Xi^{*0}\pi^0)-
\mathcal{A}(\Xi_c^0\rightarrow\Omega^-K^+)=0,
\end{equation}
\begin{equation}
\sqrt6\mathcal{A}(\Lambda_c^+\rightarrow\Sigma^{*+}\eta_8)+
\mathcal{A}(\Lambda_c^+\rightarrow\Xi^{*0}K^+)+
\sqrt6\mathcal{A}(\Xi_c^0\rightarrow \Xi^{*0}\eta_8)+
\sqrt3\mathcal{A}(\Xi_c^0\rightarrow\Omega^-K^+)=0,
\end{equation}
\begin{equation}
\sqrt2\mathcal{A}(\Lambda_c^+\rightarrow\Sigma^{*0}\pi^+)+
\mathcal{A}(\Lambda_c^+\rightarrow\Delta^+\overline{K}^0)+
\sqrt2\mathcal{A}(\Xi_c^0\rightarrow \Sigma^{*0}\overline{K}^0)+
\mathcal{A}(\Xi_c^0\rightarrow\Xi^{*-}\pi^+)=0,
\end{equation}
\begin{equation}
\sqrt{3}\mathcal{A}(\Lambda_c^+\rightarrow\Sigma^{*+}K^0)
-\mathcal{A}(\Lambda_c^+\rightarrow\Delta^{++}\pi^-)
+\sqrt{3}\mathcal{A}(\Xi_c^0\rightarrow\Xi^{*0}K^0)
-\sqrt{3}\mathcal{A}(\Xi_c^0\rightarrow\Sigma^{*+}\pi^-)=0,
\end{equation}
\begin{equation}
\sqrt{2}\mathcal{A}(\Lambda_c^+\rightarrow\Delta^0K^+)
+\sqrt{2}\mathcal{A}(\Xi_c^0\rightarrow\Sigma^{*-}K^+)
-\mathcal{A}(\Xi_c^0\rightarrow\Delta^0\pi^0)
-\sqrt3\mathcal{A}(\Xi_c^0\rightarrow\Delta^0\eta_8)=0,
\end{equation}
\begin{equation}
\sqrt{3}\mathcal{A}(\Lambda_c^+\rightarrow\Sigma^{*+}\pi^0)+
\sqrt{2}\mathcal{A}(\Lambda_c^+\rightarrow\Delta^{++}K^-)-
\sqrt{6}\mathcal{A}(\Lambda_c^+\rightarrow\Xi^{*0}K^+)+
3\mathcal{A}(\Lambda_c^+\rightarrow\Sigma^{*+}\eta_8)=0,
\end{equation}
\begin{equation}
\sqrt{3}\mathcal{A}(\Xi_c^0\rightarrow\Xi^{*0}\pi^0)+
\sqrt{6}\mathcal{A}(\Xi_c^0\rightarrow\Sigma^{*+}K^-)-
\sqrt{2}\mathcal{A}(\Xi_c^0\rightarrow\Omega^-K^+)+
3\mathcal{A}(\Xi_c^0\rightarrow\Xi^{*0}\eta_8)=0,
\end{equation}
\begin{equation}
\sqrt{6}\mathcal{A}(\Xi_c^+\rightarrow\Xi^{*0}K^+)-
\sqrt{2}\mathcal{A}(\Xi_c^+\rightarrow\Delta^{++}K^-)-
\sqrt{3}\mathcal{A}(\Xi_c^+\rightarrow\Sigma^{*+}\pi^0)-
3\mathcal{A}(\Xi_c^+\rightarrow\Sigma^{*+}\eta_8)=0,
\end{equation}
\begin{equation}
\sqrt3\mathcal{A}(\Lambda_c^+\rightarrow\Sigma^{*+}\pi^0)-
\mathcal{A}(\Lambda_c^+\rightarrow\Sigma^{*+}\eta_8)+
\sqrt3\mathcal{A}(\Xi_c^0\rightarrow \Xi^{*0}\pi^0)-
\mathcal{A}(\Xi_c^0\rightarrow\Xi^{*0}\eta_8)=0,
\end{equation}
\begin{align}
&\sqrt{2}\mathcal{A}(\Lambda_c^+\rightarrow\Delta^+\pi^0)
-2\sqrt2\mathcal{A}(\Lambda_c^+\rightarrow\Sigma^{*0}K^+)
+3\sqrt{6}\mathcal{A}(\Lambda_c^+\rightarrow\Delta^+\eta_8)\nonumber\\
&~~~~~~~~~~~+\mathcal{A}(\Xi_c^0\rightarrow\Xi^{*-}K^+)
+2\mathcal{A}(\Xi_c^0\rightarrow\Delta^+K^-)=0,
\end{align}
\begin{equation}
  \sqrt3\mathcal{A}(\Lambda_c^+\to\Sigma^{*+}\pi^0)-\mathcal{A}(\Lambda_c^+\to\Sigma^{*+}\eta_8)
  +\sqrt3\mathcal{A}(\Xi_c^0\to\Xi^{*0}\pi^0)-\mathcal{A}(\Xi_c^0\to\Xi^{*0}\eta_8)=0.
\end{equation}

\end{document}